\documentclass{article}

\usepackage{arxiv}

\usepackage[utf8]{inputenc} % allow utf-8 input
\usepackage[T1]{fontenc}    % use 8-bit T1 fonts
\usepackage{hyperref}       % hyperlinks
\usepackage{url}            % simple URL typesetting
\usepackage{booktabs}       % professional-quality tables
\usepackage{amsfonts}       % blackboard math symbols
\usepackage{nicefrac}       % compact symbols for 1/2, etc.
\usepackage{microtype}      % microtypography
\usepackage{lipsum}
\usepackage{graphicx}
\graphicspath{ {./images/} }
\usepackage{amsfonts}
\usepackage{amsmath}
\usepackage{graphics}
\usepackage{amssymb}

\usepackage[pdftex]{color}% Cores no texto

\usepackage{tabularx}
\usepackage{longtable}
\usepackage{pdflscape}
\usepackage{multirow}
\usepackage{booktabs}
\usepackage{setspace}
\usepackage[authoryear]{natbib}
\usepackage{enumitem}

\newcommand{\Cov}{\mathrm{Cov}}

\newcommand{\V}{\textbf{V}}
\newcommand{\C}{\textbf{C}}
\newcommand{\X}{\textbf{X}}
\newcommand{\Y}{\textbf{Y}}

\newcommand{\x}{\textbf{x}}
\newcommand{\w}{\textbf{w}}
\newcommand{\y}{\textbf{y}}
\newcommand{\au}{\textbf{u}}

\newcommand{\G}{\mathcal{G}}
\newcommand{\N}{\mathcal{N}}
\newcommand{\E}{\mathbb{E}}

\newcommand{\bmu}{\boldsymbol{\mu}}
\newcommand{\bSigma}{\boldsymbol{\Sigma}}
\newcommand{\bbeta}{\boldsymbol{\beta}}

\newcommand{\btheta}{\boldsymbol{\theta}}
\newcommand{\balpha}{\boldsymbol{\alpha}}
\newcommand{\bPhi}{\boldsymbol{\Phi}}
\newcommand{\bphi}{\boldsymbol{\phi}}

\newcommand{\amax}[1]{\underset{#1}{\operatorname{argmax}}}

\usepackage{color, xcolor}

\title{Censored autoregressive regression models with  \\  Student-$t$ innovations}

\author{
 Katherine A. L. Valeriano \\
  Departamento de Estatística\\
  Universidade Estadual de Campinas\\
  Campinas, Brazil 13083-970 \\
  \texttt{katandreina@gmail.com} \\
   \And
 Fernanda L. Schumacher \\
  Departamento de Estatística\\
  Universidade Estadual de Campinas\\
  Campinas, Brazil 13083-970 \\
  \texttt{fernandalschumacher@gmail.com} \\
  \And
 Christian E. Galarza \\
  Departamento de Matemáticas\\
  Escuela Superior Politécnica del Litoral\\
  Guayaquil, Ecuador 090112 \\
  \texttt{chedgala@espol.edu.ec} \\
   \And
 Larissa A. Matos \\
  Departamento de Estatística\\
  Universidade Estadual de Campinas\\
  Campinas, Brazil 13083-970 \\
  \texttt{larissam@unicamp.br} \\
}

\allowdisplaybreaks

\begin{document}
\maketitle
\begin{abstract}
The Student-$t$ distribution is widely used in statistical modeling of datasets involving outliers since its longer-than-normal tails provide a robust approach to hand such data. Furthermore, data collected over time may contain censored or missing observations, making it impossible to use standard statistical procedures. This paper proposes an algorithm to estimate the parameters of a censored linear regression model when the regression errors are autocorrelated and the innovations follow a Student-$t$ distribution.
To fit the proposed model, maximum likelihood estimates are obtained throughout the SAEM algorithm, which is a stochastic approximation of the EM algorithm useful for models in which the E-step does not have an analytic form. The methods are illustrated by the analysis of a real dataset that has left-censored and missing observations. We also conducted two simulations studies to examine the asymptotic properties of the estimates and the robustness of the model.
\end{abstract}

% keywords can be removed

\section{Introduction}

\noindent A linear regression model is a commonly used statistical tool to analyze the relation between a response (dependent) and some explanatory (independent) variables. In these models, the errors are usually considered as independent and identically distributed random variables with zero-mean and constant-variance. However, observations collected over time are often autocorrelated rather than independent. Examples of this kind of data abound in economics, medicine, environmental monitoring, and social sciences and therefore the study of the dependence among observations is of considerable practical interest \citep[see][]{tsay2005analysis, box2015time}.

A stochastic model that has been successfully used in many real-world applications to deal with serial correlation in the error term is the autoregressive (AR) model. In the AR model, the current state of the process is expressed as a finite linear combination of previous states and a stochastic shock of disturbance, which is also known as innovation in the time series literature. In general, it is assumed that the disturbance is normally distributed. For example, \cite{beach1978maximum} suggested a maximum likelihood (ML) estimation method to estimate the parameters of AR(1) error term regression model,  \cite{ullah1983estimation} used the two-stage Prais-Winsten estimator for a linear regression model with first-order autocorrelated disturbances, where a Gaussian assumption is considered for the innovations to derive a large-sample asymptotic approximation for the variance-covariance matrix, and more recently \cite{alpuim2008efficiency} proposed a linear regression model with the sequence of error terms following a Gaussian autoregressive stationary process, in which the parameters of the model are estimated using ML and ordinary least squares. 

In practice, however, the normality assumption may be unrealistic, especially in the presence of outliers. To relax this assumption, \cite{tiku1999estimating} suggested considering non-normal symmetric innovations in a simple regression model with AR(1) error term. \cite{tuacc2018robust} proposed a model with AR(p) error term considering the $t$ distribution with fixed degrees of freedom as an alternative to the normal distribution and applying the conditional maximum likelihood method to find the estimators of the model parameters, and \cite{nduka2018based} developed an EM algorithm to estimate the parameters in autoregressive models of order $p$ with Student-$t$ innovations.

An additional challenge arises when some observations are censored or missing. In the first case, values can occur out of the range of a measuring instrument, and in the second, the value is unknown. Censored time series data are frequently encountered in environmental monitoring, medicine, and economics, and there are some proposals in the literature related to censored autoregressive linear regression models with normal innovations. For example, \cite{zeger1986regression} proposed a Gaussian censored autocorrelated model with parameters estimated by maximizing the full likelihood using the quasi-Newton and Expectation-Maximization (EM) algorithms, and the authors pointed out that the quasi-Newton approach is preferable because it estimates the variance-covariance matrix for the parameters. 

Additionally, \cite{park2007censored} developed a censored autoregressive moving average model with Gaussian white noise, in which the algorithm works iteratively imputing the censored or missing observations by sampling from the truncated normal distribution and the parameters are estimated by maximizing an approximate full likelihood function, and \cite{wang2018quasi} suggested a quasi-likelihood method using the complete-incomplete data framework. Recently, \cite{schumacher2017censored} considered an autoregressive censored linear model with the parameters estimated via an analytically tractable and efficient stochastic approximation of the EM (SAEM) algorithm, and \cite{liu2019parameter} proposed a coupled Monte Carlo Markov Chain (MCMC)-SAEM algorithm to fit an AR$(p)$ regression model with Student-$t$ innovations accounting for missing data.

Nevertheless, to the best of our knowledge there are no studies that consider the Student-$t$ distribution for the innovations in censored autoregressive models from a likelihood-based perspective. Thence, in this work we propose an EM-type algorithm to estimate the parameters of a censored regression model with autoregressive errors and innovations following a Student-$t$ distribution. Specifically, the SAEM algorithm is considered to avoid the direct computation of complex expressions on the E-step of the EM algorithm \citep{dempster1977maximum} and since its computational effort is much smaller in comparison to the Monte Carlo EM algorithm \citep{wei1990monte}, as shown by \cite{jank2006implementing}.

The paper is organized as follows. Section \ref{preliminaries} introduces the autoregressive regression models of order $p$ and the SAEM algorithm. Section \ref{modelAR} is devoted to formulate the censored autoregressive model with Student-$t$ innovations, providing the expressions used to estimate the parameters of the proposed model, the method used to estimate the observed information matrix, and the expression to make predictions. Section \ref{simulation} displays some results of two simulation studies carried out to examine the asymptotic properties of the estimators, and to demonstrate the robustness of the model. Section \ref{application} applies the proposed model to the total phosphorus concentration dataset available in the \textsf{R} package \textbf{ARCensReg} \citep[][]{schumacher2016package, rteam}, along with an analysis of quantile residuals. Finally, Section \ref{conclusions} concludes with a discussion.

%%%%%%%%%%%%%%%%%%%%%%%%%%%%%%%%%%%%%%% 

% https://onlinelibrary.wiley.com/doi/epdf/10.1111/1467-9574.00055
% artigo sobre problema de usar como uma realização multivariada 

\section{Preliminaries}\label{preliminaries}

We begin by defining some notations and introducing the basic definitions and algorithms used throughout this work. The normal distribution with mean $\mu$ and variance $\sigma^2$ is denoted by $\N(\mu,\sigma^2)$, the notation for a $p$-variate normal distribution with mean $\bmu$ and variance-covariance matrix $\bSigma$ is $\N_p(\bmu,\bSigma)$, and a truncated multivariate normal distribution with truncation region $\mathbb{A}$ is denoted by $\mbox{TN}_p(\bmu,\bSigma; \mathbb{A})$. The notation for a random variable with Gamma distribution is given by $\mathcal{G}(\alpha,\beta)$, where $\alpha>0$ is the shape parameter and $\beta>0$ is the rate parameter. The Student-$t$ distribution with location parameter $\mu$, scale parameter $\sigma^2>0$, and $\nu>0$ degrees of freedom is denoted by $t(\mu,\sigma^2,\nu)$. A random variable $X$ with uniform distribution on the interval $(a,b)$ is denoted by $X\sim \mathcal{U}(a,b)$. Furthermore, the expression ``$iid$'' means independent and identically distributed, ${\bf A}^\top$ represents the transpose of ${\bf A}$, $\Gamma(a)=\int_0^\infty x^{a-1}e^{-x}\mathrm{d}x$ is the gamma function evaluated at $a>0$, and $\mathcal{F}_t = \sigma(Y_1,Y_2,\ldots, Y_t)$ is the $\sigma$-field generated by $\{Y_1,Y_2,\ldots, Y_t\}$. Finally, for integers $2\leq t \leq n$ and $1 \leq p < t$, we use the index $(t,p)$ as follows: for a vector $\y = (y_1, y_2, \ldots, y_n)^\top \in \mathbb{R}^n$, then $\y_{(t,p)} = (y_{t-1}, y_{t-2}, \ldots, y_{t-p})^\top \in \mathbb{R}^p$; and for a matrix $\X \in \mathbb{R}^{n \times q}$ with rows $\x_i \in \mathbb{R}^q$, $1\leq i \leq n$, then $\X_{(t,p)} = [\x_{t-1}, \x_{t-2}, \ldots, \x_{t-p}]^\top \in \mathbb{R}^{p\times q}$.

\subsection{The autoregressive regression model of order $p$} \label{SMN}

First, consider a linear regression model with errors that are autocorrelated as a discrete-time autoregressive process of order $p$; thus, the model for an observation at time $t$ is given by
\begin{eqnarray}
Y_t &=& \x_t^\top\bbeta + \xi_t, \label{modelo1} \\
\xi_t &=& \phi_1 \xi_{t-1} + \ldots + \phi_p \xi_{t-p} + \eta_t, \quad \eta_t\stackrel{iid}{\sim} F(\cdot), \quad t = 1,\ldots,n, \label{error1}
\end{eqnarray}
where $Y_t$ represents the dependent variable observed at time $t$, $\x_t=(x_{t1}, \ldots, x_{tq})^\top$ is a $q\times 1$ vector of known covariables, $\bbeta$ is a $q\times 1$ vector of unknown regression parameters to be estimated, and $\xi_{t}$ is the regression error and is assumed to follow an autoregressive model, with $\bphi = (\phi_1, \ldots, \phi_p)^\top$ begin a $p\times 1$ vector of autoregressive coefficients and $\eta_t$ being a shock of disturbance with distribution $F(\cdot)$. The term denoted by $\eta_t$ is also known as the innovation in the time series literature \citep[see, for instance,][]{box2015time, schumacher2017censored, wang2018quasi}.

Now, suppose that $\eta_t$ in Equation \ref{error1} follows a Student-$t$ distribution with location parameter 0, scale parameter $\sigma^2 > 0$, and $\nu > 0$ degrees of freedom, whose probability density function (pdf) can be written as
\begin{equation}\label{tdistribution}
f(\eta; \sigma^2, \nu) = \frac{\Gamma\left(\frac{\nu+1}{2}\right)}{\Gamma\left(\frac{\nu}{2}\right)(\pi\nu\sigma^2)^{\frac{1}{2}}} \left( 1 + \frac{\eta^2}{\nu\sigma^2}\right)^{-\frac{\nu + 1}{2}}, \quad \eta\in\mathbb{R}.
\end{equation}
Then, the model defined by Equations \ref{modelo1}-\ref{tdistribution} will be called the {\it autoregressive regression $t$ model of order $p$} (AR$t(p)$) hereinafter. 

Particularly, the distribution of $\eta_t\sim t(0,\sigma^2,\nu)$ might be written using a scale mixture of normal (SMN) distribution as described in \cite{kotz2004multivariate}, where $\eta_t$ is equal in distribution to $Z_t/\sqrt{U_t}$, with $Z_t\stackrel{iid}{\sim}\N(0,\sigma^2)$,  $U_t\stackrel{iid}{\sim}\G(\nu/2,\nu/2)$, and $Z_t$ being independent of $U_t$. Consequently, the conditional distribution of $\eta_t$ given $U_t=u_t$ is normal with zero mean and variance $\sigma^2/u_t$, i.e.,
$$\eta_t| \, (U_t=u_t) \stackrel{ind}{\sim} \N(0,\sigma^2/u_t) \quad \mbox{and} \quad U_t\stackrel{iid}{\sim}\mathcal{G}(\nu/2, \nu/2), \quad t=1,\ldots,n.$$ 
This representation facilitates the implementation of an EM-type algorithm, which is discussed next.

\subsection{The SAEM algorithm}\label{saem}

The EM algorithm was first introduced by \cite{dempster1977maximum} and provides a general approach to the iterative computation of ML estimates in problems with incomplete data. Let $\y_c=(\y_o^\top, \y_m^\top)^\top$ be the complete data, where $\y_o$ is the observed data and $\y_m$ is the incomplete (or missing) data. Let $\ell_c(\btheta; \y_c)$ be the complete log-likelihood function. The EM algorithm proceeds as follows:

\begin{itemize}[leftmargin=0.5cm]
\item \textbf{E-step}: Let $\widehat{\btheta}^{(k)}$ be the estimate of $\btheta$ at the $k$th iteration. Compute the conditional expectation of the complete log-likelihood function $Q_{k}(\btheta) = \E\left[\ell_c(\btheta; \y_c)|\y_o,\widehat{\btheta}^{(k)}\right]$.
\item \textbf{M-step}: Maximize $Q_{k}(\btheta)$ with respect to $\btheta$ to obtain $\widehat{\btheta}^{(k+1)}$.
\end{itemize}

There are situations where the E-step has no analytic form. To deal with these cases, \cite{wei1990monte} proposed to replace the expectation in the computation of $Q_k(\btheta)$ with a Monte Carlo (MC) approximation based on a large number of independent simulations of the missing data, which was called the Monte Carlo EM (MCEM) algorithm. As an alternative to the computationally intensive MCEM algorithm, \cite{delyon1999convergence} proposed a scheme that splits the E-step into a simulation step and an integration step (using a stochastic approximation), while the maximization step remains unchanged. This algorithm was called the SAEM algorithm and performs as follows:

\begin{itemize}[leftmargin=0.5cm]
\item {\bf E-step}:
\begin{itemize}
\item[1.] Simulation: Draw $M$ samples of the missing data $\y_m^{(l,k)}, l=1,\ldots, M$ from the conditional distribution $f(\y_m; \widehat{\btheta}^{(k)}, \y_o)$. 

%\hspace*{1.4cm}
\item[2.] Stochastic approximation: Update ${Q}_k(\btheta)$ by 
\begin{eqnarray*}
\widehat{Q}_k(\btheta) = \widehat{Q}_{k-1}(\btheta) + \delta_k \left(\frac{1}{M}\sum_{l=1}^M \ell_c(\btheta; \y_m^{(k,l)},\y_o) - \widehat{Q}_{k-1}(\btheta)\right),
\end{eqnarray*}
where $\delta_k$ is a decreasing sequence of positive numbers such that $\sum_{k=1}^\infty \delta_k = \infty$ and $\sum_{k=1}^\infty \delta_k^2 < \infty$, also known as the smoothness parameter \citep{kuhn2005maximum}.

\end{itemize}

\item \textbf{M-step}: Update $\widehat{\btheta}^{(k)}$ as $\widehat{\btheta}^{(k+1)}=\amax{\btheta} \;\widehat{Q}_k(\btheta)$.
\end{itemize}
This process is iterated until some distance between two successive evaluations of the actual log-likelihood function becomes small enough.

%%%%%%%%%%%%%%%%%%%%%%%%%%%%%%%%%%%%%%% 
\section{Proposed model and parameter estimation}\label{modelAR}

This section is devoted the formulating and the ML estimation of the censored autoregressive linear model. Here, we specify the log-likelihood function, the algorithm used to overcome the parameter estimation problem, expressions to approximate the standard errors, and expressions to predict at unobserved times when the values of some covariables (if needed) at these times are available.

\subsection{The censored ART($p$) model}

Assume that the response variable $Y_t$ given in Equation \ref{modelo1} is not fully observed for all times. Instead, we observe $(V_t, C_t)$, where $V_t$ represents either an observed value or the limit of detection (LOD) of a censored variable, and $C_t$ is the censoring indicator defined as
\begin{eqnarray}\label{censored}
C_t = \left\lbrace \begin{array}{lll}
1 & \mbox{if } \quad V_{t1} \leq Y_t \leq V_{t2}, & (\mbox{censored}) \\
0 & \mbox{if } \quad V_t = Y_t. & (\mbox{observed})
\end{array}\right.
\end{eqnarray}
Thereby, the model defined by Equations \ref{modelo1}-\ref{censored} will be called {\it censored autoregressive regression $t$ model of order $p$} (CAR$t(p)$). We say that $Y_t$ is left-censored if $C_t=1$ and $V_t=(-\infty,V_{t2}]$, right-censored if $C_t=1$ and $V_t=[V_{t1},+\infty)$, interval censored if $C_t=1$ and $V_t=[V_{t1},V_{t2}]$, and it represents a missing value if $C_t=1$ and $V_t=(-\infty,+\infty)$.

To compute the log-likelihood function of the CAR$t(p)$ model, we condition the marginal distribution on the first $p$ observations, which are considered fully observed, and we denote by $\y_o \in \mathbb{R}^{n_o}$ the vector of observed data and by $\y_m\in \mathbb{R}^{n_m}$ the vector of censored or missing observations, with $n_o + n_m = n-p$. Note also that the distribution of $Y_t$ conditional to all the preceding data $\mathcal{F}_{t-1}$ only depends on the previous $p$ observations. Then, for $\btheta=(\bbeta^\top, \bphi^\top, \sigma^2, \nu)^\top$, $Y_t$ given $\y_{(t,p)}$ follows a Student-$t$ distribution with location parameter $\mu_t=\x_t^\top\bbeta + (\y_{(t,p)} - \X_{(t,p)}\bbeta)^\top\bphi$, scale parameter $\sigma^2$, and $\nu$ degrees of freedom, where $\X_{(t,p)}$ is a $p\times q$ matrix with the covariates related to $\y_{(t,p)}$ and $\x_t$ is a vector with the covariates related to $Y_t$. In other words, we have %, i.e., 
\begin{equation*}
f(y_t|\btheta,\mathcal{F}_{t-1}) = f(y_t|\btheta,\y_{(t,p)}) = f(y_t|\btheta, y_{t-1}, y_{t-2}, \ldots, y_{t-p}), \quad t=p+1,\ldots, n.
\end{equation*}
Thus, the observed (conditional) log-likelihood function can be computed by
\begin{equation}\label{likelio}
\ell(\btheta; \y_o) = \log\left( \int_{\mathbb{R}^m} \prod_{t=p+1}^n f(y_t|\btheta, y_{t-1}, y_{t-2}, \ldots, y_{t-p}) \mathrm{d}\y_m \right).
\end{equation}

%%%%%%%%%%%%%%%%%%%%%%%%%%%%%%%%%%%%%%% 

\subsection{Parameter estimation via the SAEM algorithm}

The observed log-likelihood function in Equation \ref{likelio} involves complex expressions, making work directly with $\ell(\btheta; \y_o)$ very difficult. Hence, to obtain the ML estimates of $\btheta$, we now apply an EM-type algorithm considering the hierarchical representation %in the SMN distributions mentioned 
presented in Subsection \ref{SMN}. 

Let $\y=(y_{p+1},\ldots,y_n)^\top$ and $\au=(u_{p+1},\ldots,u_n)^\top$ be hypothetical missing data, and consider that we observe $(\V,\C)$, where $\V = (V_{p+1}, \ldots, V_n)^\top$ and $\C = (C_{p+1}, \ldots, C_n)^\top$. Then, the complete dataset is given by $\y_c = \left(\y^\top, \au^\top, \V^\top, \C^\top\right)^\top$, and the complete log-likelihood function $\ell_{c}(\btheta; \y_c)$ can be written as \begin{eqnarray*}
\ell_c(\btheta; \y_c) &=& g(\nu| \au) - \frac{n-p}{2} \log\sigma^2 - \frac{1}{2\sigma^2}\sum_{i=p+1}^n u_i\left(\tilde{y}_i - \w_{i}^\top\bphi\right)^2 + \textrm{cte},
\end{eqnarray*}
with $g(\nu| \au) = \frac{n-p}{2} \left[ \nu\log\left(\frac{\nu}{2}\right) - 2\log\Gamma\left(\frac{\nu}{2}\right) \right] + \frac{\nu}{2} \left(\sum_{i=p+1}^n\log u_i - \sum_{i=p+1}^n u_i\right)$, $\tilde{y}_t = y_t - \x_t^\top\bbeta$, $\w_{t} = \y_{(t,p)} - \X_{(t,p)}\bbeta$, and $\textrm{cte}$ being a constant.

Let $\widehat{\btheta}^{(k)}$ denote the current estimate of $\btheta$. Then, the conditional expectation of the complete log-likelihood function given the observed data and ignoring constant terms is given by
\begin{eqnarray*}
Q_{k}(\btheta) &=& \E\left[\ell_c(\btheta; \y_c)|\V,\C,\widehat{\btheta}^{(k)}\right] \\
&=& \widehat{g(\nu| \au)}^{(k)} - \frac{n-p}{2}\log\widehat{\sigma}^{2(k)} - \frac{1}{2\widehat{\sigma}^{2(k)}} \left( \widehat{u y^2_*}^{(k)} - 2 \widehat{\bphi}^{(k)\top} \widehat{u y\y}_*^{(k)} + \widehat{\bphi}^{(k)\top}\widehat{u\y^2}_*^{(k)}\widehat{\bphi}^{(k)} \right), \hspace*{1cm} \label{Qfun}
\end{eqnarray*}
where 
\begin{eqnarray*}
\widehat{g(\nu| \au)}^{(k)} &=& \frac{n-p}{2} \left[ \widehat{\nu}^{(k)}\log\left(\frac{\widehat{\nu}^{(k)}}{2}\right) - 2\log\Gamma\left(\frac{\widehat{\nu}^{(k)}}{2}\right) \right] + \frac{\widehat{\nu}^{(k)}}{2} \left(\widehat{\mbox{log(u)}}^{(k)} - \widehat{u}^{(k)}\right), \\
\widehat{u y^2_{*}}^{(k)} &=& \widehat{u y^2}^{(k)} - \sum_{i=p+1}^n \left( 2 \widehat{u y_i}^{(k)}\x_i^\top\widehat{\bbeta}^{(k)} -  \widehat{u}_i^{(k)}\widehat{\bbeta}^{(k)\top}\x_i\x_i^\top\widehat{\bbeta}^{(k)} \right), \\ 
\widehat{uy\y_*}^{(k)} &=& \widehat{uy\y}^{(k)} - \sum_{i=p+1}^n \left( \widehat{u y_i}^{(k)}\X_{(i,p)}\widehat{\bbeta}^{(k)} + \widehat{u\y_i}^{(k)} \x_i^\top\widehat{\bbeta}^{(k)} - \widehat{u}_i^{(k)} \X_{(i,p)}\widehat{\bbeta}^{(k)}\widehat{\bbeta}^{(k)\top}\x_i\right), \\
\widehat{u\y^2_*}^{(k)} &=& \widehat{u\y^2}^{(k)} - \sum_{i=p+1}^n\left( \widehat{u\y_i}^{(k)}\widehat{\bbeta}^{(k)\top}\X_{(i,p)}^\top + \X_{(i,p)}\widehat{\bbeta}^{(k)}\widehat{u\y_i}^{(k)\top} - \widehat{u}_i^{(k)} \X_{(i,p)}\widehat{\bbeta}^{(k)}\widehat{\bbeta}^{(k)\top}\X_{(i,p)}^\top\right),
\end{eqnarray*}
such that 
$\widehat{u}^{(k)} = \sum_{i=p+1}^n \widehat{u}_i^{(k)}$, $\widehat{u}_i^{(k)} = \E[ U_i|\V,\C,\widehat{\btheta}^{(k)} ]$, $\widehat{\mbox{log(u)}}^{(k)} = \E[\sum_{i=p+1}^n\log (U_i)|\V,\C,\widehat{\btheta}^{(k)} ]$,
$\widehat{u y^2}^{(k)} = \E[\sum_{i=p+1}^n U_iY_i^2| \V,\C,\widehat{\btheta}^{(k)}]$, 
$\widehat{uy_i}^{(k)} = \E[U_i Y_i|\V,\C,\widehat{\btheta}^{(k)}]$, 
$\widehat{u y\y}^{(k)} = \E[ \sum_{i=p+1}^n U_iY_i\Y_{(i,p)}|\V,\C,\widehat{\btheta}^{(k)}]$, 
$\widehat{u\y_i}^{(k)} = \E[ U_i\Y_{(i,p)}|\V,\C,\widehat{\btheta}^{(k)}]$, and $\widehat{u\y^2}^{(k)} = \E[\sum_{i=p+1}^n U_i\Y_{(i,p)}\Y_{(i,p)}^\top|\V,\C,\widehat{\btheta}^{(k-1)}]$, for $i= p+1, \ldots, n$. 

\indent It is easy to observe that the E-step reduces to the computation of $\widehat{u}_i^{(k)}$, $\widehat{\mbox{log(u)}}^{(k)}$, $\widehat{u y^2}^{(k)}$, $\widehat{u_i\y_i}^{(k)}$, $\widehat{u y\y}^{(k)}$, $\widehat{u\y_i}^{(k)}$, and $\widehat{u\y^2}^{(k)}$, for $i=p+1,\ldots,n$. Because of the difficulties in the calculation of the conditional expectations in the E-step, specifically, when we have successive censored observations, we consider the implementation of the SAEM algorithm, introduced in Subsection \ref{saem}, which at the $k$th iteration proceeds as follows: \\

\noindent \textbf{Step E-1 (Sampling)}. We first consider the observed and censored/missing part of $\y = (\y_o^\top, \y_m^\top)^\top$ separately and rearrange the elements of the location vector and scale matrix parameters to compute the conditional distribution of $\y_m$ as demonstrated in Appendix \ref{conditional}. Thus, in the simulation step, we generate $M$ samples from the conditional distribution of the latent variables $(\y,\au)$ via the Gibbs sampler algorithm according to the following scheme: 

\begin{itemize}[leftmargin=0.5cm]
\item \textbf{Step 1}: Sample $\y_m^{(k,l)}$ from the $\mbox{TN}_{n_m}(\tilde{\bmu}^{*(k)}, \tilde{\bSigma}^{*(k)}; \V_m)$, with $\tilde{\bmu}^{*(k)} = \tilde{\bmu}_m^{(k)} + \tilde{\bSigma}_{mo}^{(k)}\tilde{\bSigma}_{oo}^{-1(k)}(\y_o - \tilde{\bmu}_o^{(k)})$, $\tilde{\bSigma}^{*(k)} = \tilde{\bSigma}_{mm}^{(k)} - \tilde{\bSigma}_{mo}^{(k)}\tilde{\bSigma}_{oo}^{-1(k)}\tilde{\bSigma}_{om}^{(k)}$, and $\textbf{V}_m = \{ \y_m=(y_1^m, \ldots, y_{n_m}^m)^\top; V_{11}^m \leq y_1^m \leq V_{12}^m, \ldots, V_{n_m1}^m \leq \y_{n_m}^m \leq V_{n_m2}^m \}$, where the parameters $\tilde{\bmu}_m^{(k)}, \tilde{\bmu}_o^{(k)}, \tilde{\bSigma}_{mm}^{(k)}, \tilde{\bSigma}_{mo}^{(k)}$, and $\tilde{\bSigma}_{oo}^{(k)}$ are computed using Equations \ref{media1} and \ref{var1} available in Appendix \ref{conditional}. Then, we construct a full vector of observations as $\y^{(k,l)} = (\y_{o}^{\top}, \y_m^{(k,l)\top})^\top$, using the observed  and the sample generated values, for $l = 1, \ldots, M$.

\item \textbf{Step 2}: Sample $\au^{(k,l)}$ from $f(\au|\y^{(k,l)},\widehat{\btheta}^{(k)})$, whose elements are independent and Gamma distributed as $u_i^{(k,l)}|\y^{(k,l)},\widehat{\btheta}^{(k)}\stackrel{ind}{\sim}\G(a_i^{(k,l)}, b_i^{(k,l)})$ with $a_i^{(k,l)} = (\widehat{\nu}^{(k)} + 1)/2$, $b_i^{(k,l)} = (\widehat{\nu}^{(k)} + \varrho_i^{(k,l)2}/\widehat{\sigma}^{2(k)} )/2$, and $\varrho_i^{(k,l)} = y_i^{(k,l)} - \x_i^\top\widehat{\bbeta}^{(k)} - \y_{(i,p)}^{(k,l)\top}\widehat{\bphi}^{(k)} + \widehat{\bbeta}^{(k)\top}\X_{(i,p)}^\top\widehat{\bphi}^{(k)}$, for $i=p+1, \ldots, n$ and $l = 1, \ldots, M$.
\end{itemize}

\noindent \textbf{Step E-2 (Stochastic Approximation)}. Given the sequence $(\y^{(k,l)}, \au^{(k,l)})$ for $l = 1,\ldots,M$, we replace the conditional expectations in $Q_k(\btheta)$ by the following stochastic approximations:
\begin{eqnarray*}
&& \widehat{u}_i^{(k)} = \widehat{u}_i^{(k-1)} + \delta_k \left( \frac{1}{M} \sum_{l=1}^M u_i^{(k,l)} - \widehat{u}_i^{(k-1)} \right), \quad i=p+1,\ldots,n. \\
&& \widehat{\mbox{log(u)}}^{(k)} = \widehat{\mbox{log(u)}}^{(k-1)} + \delta_k \left(\frac{1}{M} \sum_{l=1}^M \sum_{i=p+1}^n \log u_i^{(k,l)} - \widehat{\mbox{log(u)}}^{(k-1)}\right). \\
&& \widehat{u y^2}^{(k)} = \widehat{u y^2}^{(k-1)} + \delta_k \left(\frac{1}{M} \sum_{l=1}^M \sum_{i=p+1}^n u_i^{(k,l)}y_i^{2(k,l)} - \widehat{u y^2}^{(k-1)}\right). \\
&& \widehat{uy_i}^{(k)} = \widehat{uy_i}^{(k-1)} + \delta_k \left(\frac{1}{M} \sum_{l=1}^M  u_i^{(k,l)}y_i^{(k,l)} - \widehat{uy_i}^{(k-1)}\right), \quad i=p+1,\ldots, n. \\
&& \widehat{u y\y}^{(k)} = \widehat{u y\y}^{(k-1)} + \delta_k \left(\frac{1}{M} \sum_{l=1}^M \sum_{i=p+1}^n u_i^{(k,l)}y_i^{(k,l)}\y_{(i,p)}^{(k,l)} - \widehat{u y\y}^{(k-1)}\right). \\
&& \widehat{u\y_i}^{(k)} = \widehat{u\y_i}^{(k-1)} + \delta_k \left(\frac{1}{M} \sum_{l=1}^M  u_i^{(k,l)}\y_{(i,p)}^{(k,l)} - \widehat{u\y_i}^{(k-1)}\right), \quad i=p+1,\ldots,n. \\
&& \widehat{u\y^2}^{(k)} = \widehat{u\y^2}^{(k-1)} + \delta_k \left(\frac{1}{M} \sum_{l=1}^M \sum_{i=p+1}^n u_i^{(k,l)}\y_{(i,p)}^{(k,l)}\y_{(i,p)}^{(k,l)\top} - \widehat{u\y^2}^{(k-1)}\right). \hspace*{5.5cm}
\end{eqnarray*}

Following \cite{galarza2017quantile}, the variable $\delta_k$ will be considered as $\delta_k=1$, if $1\leq k \leq cW$, and $\delta_k=1/(k - cW)$, if $cW + 1 \leq k \leq W$, where $W$ is the maximum number of iterations and $c$ is a cutoff point $(0 \leq c \leq 1)$ which determines the percentage of initial iterations with no-memory. If $c = 0$, the algorithm will have a memory for all iterations and hence will converge slowly to the ML estimates, and $W$ needs to be large. If $c = 1$, the algorithm will be memory-free, it will converge quickly to a solution neighborhood, and the algorithm will initiate a Markov chain leading to a reasonably well-estimated mean after applying the necessary \textit{burn-in} and \textit{thinning} steps. A number between 0 and 1 $(0 < c < 1)$ will assure an initial convergence in distribution to a solution neighborhood for the first $cW$ iterations and an almost sure convergence for the rest of the iterations. 

The selection of $c$ and $W$ could affect the speed of convergence of the SAEM algorithm. A graphical approach can monitor the convergence of the estimates for all parameters and determine the values for these constants, as suggested by \cite{lavielle2014mixed}. An advantage of the SAEM algorithm is that, even though it performs an MCMC E-step, it only requires a small and fixed sample size $M$ %$N$, 
making it much faster than the MCEM algorithm. \\

\noindent \textbf{Step CM (Conditional Maximization)}. A conditional maximization step is then carried out, and $\widehat{\btheta}^{(k)}$ is updated by maximizing $Q_k(\btheta)$ over $\btheta$ to obtain a new estimate $\widehat{\btheta}^{(k+1)}$, leading to the expressions:
\begin{eqnarray}
&& \widehat{\bphi}^{(k+1)} = \left( \widehat{u\y^2_*}^{(k)} \right)^{-1} \widehat{u y\y_*}^{(k)}, \\
&& \widehat{\sigma}^{2(k+1)} = \frac{1}{n-p} \left( \widehat{u y_*^2}^{(k)} - 2 \widehat{\bphi}^{(k+1)\top} \widehat{u y\y_*}^{(k)} + \widehat{\bphi}^{(k+1)\top}\widehat{u\y^2_*}^{(k)}\widehat{\bphi}^{(k+1)} \right), \\
&& \widehat{\bbeta}^{(k+1)} = \left(\sum_{i=p+1}^n \widehat{u}_i^{(k)} \widehat{\balpha}_i^{(k+1)}\widehat{\balpha}_i^{(k+1)\top}\right)^{-1} \sum_{i=p+1}^n \left( \widehat{u y_i}^{(k)} - \widehat{\bphi}^{(k+1)\top}\widehat{u\y_i}^{(k)} \right)\widehat{\balpha}_i^{(k+1)}, \\
&& \widehat{\nu}^{(k+1)} = \amax{ \nu }  \quad\widehat{g(\nu| u)}^{(k)}, \hspace*{11cm}
\end{eqnarray}
with $\widehat{\balpha}_i^{(k+1)} = \x_i - \X_{(i,p)}^\top\widehat{\bphi}^{(k+1)}$ for $i=p+1,\ldots,n$.

%%%%%%%%%%%%%%%%%%%%%%%%%%%%%%%%%%%%%

\subsection{Standard error approximation}
The Fisher information matrix is a good measure of the amount of information that a sample dataset provides about the parameters, and it can be used to compute the asymptotic variance of the estimators. \cite{louis1982finding} developed a procedure for extracting the observed information matrix when the EM algorithm is used to find the ML estimates in problems with incomplete data. 

Let $\mathbf{S}_c(\y_c; \btheta)$ and $\mathbf{B}_c(\y_c; \btheta)$ respectively denote the first derivative and the negative of the second derivative of the complete-data log-likelihood 
function with respect to the parameter vector $\btheta$. %of the complete data. 
Let $\mathbf{S}_o(\y_o; \btheta)$ and $\mathbf{B}_o(\y_o; \btheta)$ be the corresponding derivatives of the log-likelihood function of the observed data. Then, the observed information matrix is given by
\begin{equation}
\textbf{I}_o(\btheta) = \mathbf{B}_o(\y_o;\btheta) = \E\left[\mathbf{B}_c(\y_c;\btheta) \big|\y_o\right] - \E\left[\mathbf{S}_c(\y_c;\btheta) \mathbf{S}_c^\top(\y_c;\btheta)\big|\y_o\right] + \mathbf{S}_o(\y_o;\btheta)\mathbf{S}_o^\top(\y_o;\btheta).
\end{equation}
\cite{delyon1999convergence} adapted the method proposed by \cite{louis1982finding} to compute the observed information matrix when the SAEM algorithm is used to estimate the parameters. For this, it is necessary to compute the auxiliary terms
\begin{eqnarray}
\mathbf{H}_k &=& -\mathbf{G}_k + \mathbf{\Delta}_k  \mathbf{\Delta}_k^\top, \nonumber \\
\mathbf{G}_k &=& \mathbf{G}_{k-1} + \delta_k \left( \frac{1}{M} \sum_{l=1}^M \left( \frac{\partial^2 \ell_c(\btheta; \y^{(k,l)})}{\partial\btheta \partial \btheta^\top} + \frac{\partial\ell_c(\btheta; \y^{(k,l)})}{\partial\btheta} \frac{\partial\ell_c(\btheta; \y^{(k,l)})}{\partial\btheta}^\top \right) - \mathbf{G}_{k-1} \right), \,\,\textrm{ and } \label{IMSE} \\
\mathbf{\Delta}_k &=& \mathbf{\Delta}_{k-1} + \delta_k \left( \frac{1}{M}\sum_{l=1}^M \frac{\partial\ell_c(\btheta; \y^{(k,l)})}{\partial\btheta} - \mathbf{\Delta}_{k-1} \right), \nonumber \hspace*{8cm}
\end{eqnarray}
where $\y^{(k,l)}=(\y_o^\top,\y_m^{(k,l)\top})^\top$ is the vector of observations sampled at iteration $(k,l)$ for $k=1,\ldots,W$ and $l=1,\ldots, M$, with $W$ denoting the maximum number of iteration of the SAEM algorithm and $M$ the number of MC samples for stochastic approximation. The inverse of the limiting value of $\mathbf{H}_k$ can be used to assess the dispersion of the estimators \citep{delyon1999convergence}. The analytical expressions for the first and second derivatives of the complete data log-likelihood function are given in Appendix \ref{derivadas}.   

%%%%%%%%%%%%%%%%%%%%%%%%%%%%%%%%%%%%%%% 

\subsection{Prediction}\label{sec:prediction}
Considering the interest of predicting values from the CAR$t(p)$ model, we denote by $\y_\textrm{obs}$ the $n$-vector of random variables (RVs) corresponding to the given sample and by $\y_\textrm{pred}$ the vector of RVs of length $n_\textrm{pred}$ corresponding to the time points that we are interested in predicting.

Let $\y_\textrm{obs}=(\y_o^{\top}, \y_m^{\top})^\top$, where $\y_{o}$ is the vector of uncensored observations and $\y_{m}$ is the vector of censored or missing observations. To deal with the censored values existing in $\y_\textrm{obs}$, we use an imputation procedure that consists in replacing the censored values with the values obtained in the last iteration of the SAEM algorithm, i.e., $\y_m = \E[\y_m | \V, \C, \widehat{\btheta}^{(W)}] \approx \widehat{\y}_m^{(W)}$, since elements of $\widehat{\y}_m^{(k)}$ can also be updated during the Step E-2 of the SAEM algorithm as
\begin{equation}\label{imputed}
\widehat{\y}_m^{(k)} = \widehat{\y}_m^{(k-1)} + \delta_k \left( \frac{1}{M} \sum_{l=1}^M \y_m^{(k,l)} - \widehat{\y}_m^{(k-1)}\right), \quad k = 1, \ldots, W,
\end{equation}
with the same $\delta_k$, $M$, and $W$ settings considering in the estimation. The new vector of observed values will be denoted by $\y_\textrm{obs*} = (\y_o^\top, \widehat{\y}_m^{(W)\top})^\top$. 

Now, supposing that all values in $\y_\textrm{obs*}$ are completely observed and that the explanatory variables for $\y_\textrm{pred}$ are available, the forecasting procedure will be performed recursively \citep{box2015time} as follows
$$
\widehat{y}_{n+k} = 
\begin{cases} 
\x_{n+k}^\top\bbeta + \sum_{j=k}^p \phi_j(y_{n+k-j} - \x_{n+k-j}^\top\bbeta), & k = 1 \\
& \\
\x_{n+k}^\top\bbeta + \sum_{i=1}^{k-1} \phi_i(\widehat{y}_{n+k-i} - \x_{n+k-i}^\top\bbeta) + \sum_{j=k}^p \phi_j(y_{n+k-j} - \x_{n+k-j}^\top\bbeta), & 1 < k \leq p \\
&  \\
\x_{n+k}^\top\bbeta + \sum_{j=1}^p \phi_j(\widehat{y}_{n+k-j} - \x_{n+k-j}^\top\bbeta), & p < k \leq n_\textrm{pred}.
\end{cases}
$$
In practice, $\btheta$ is substituted by $\widehat{\btheta}=\left( \widehat{\bbeta}^\top, \widehat{\bphi}^\top, \widehat{\sigma}^2, \widehat{\nu}\right)^\top$, where $\widehat{\btheta}$ is the estimate obtained via the SAEM algorithm, such that $\widehat{\y}_\textrm{pred} = (\widehat{y}_{n+1}, \widehat{y}_{n+2}, \ldots, \widehat{y}_{n+n_\textrm{pred}})^\top$.

%%%%%%%%%%%%%%%%%%%%%%%%%%%%%%%%%%%%%%

\section{Simulation Study}\label{simulation}

In this section, we examine the asymptotic properties of the SAEM estimates through a simulation study considering different sample sizes and levels of censoring. A second simulation study is performed to demonstrate the robustness of the estimates obtained from the proposed model when the data is perturbed.

\begin{table}
\caption{\textbf{Simulation study 1}. Summary statistics of parameter estimates for the CAR$t$(2) model based on 300 samples of sizes $n=100, 300, 600$, and different levels of censoring.}\label{estimates1}
\centering
\small
\begin{tabular*}{\textwidth}{c@{\extracolsep{\fill}}cclccccccc}
\toprule
&& Average & & \multicolumn{7}{c}{Parameter} \\
\cline{5-11}
$n$ & LOD & level of & Measure & $\beta_0$ & $\beta_1$ & $\beta_2$ & $\phi_1$ & $\phi_2$ & $\sigma^2$ & $\nu$ \\
\cline{5-11}
&& censoring & & 5.00 & 0.50 & 0.90 & -0.40 & 0.12 & 2.00 & 4.00 \\ 
\midrule
\multirow{16}{*}{100} & \multirow{4}{*}{No} & \multirow{4}{*}{0\%} & MC-Mean & 4.988 & 0.501 & 0.937 & -0.409 & 0.093 & 1.779 & 4.150 \\ 
&&& IM-SE & 0.325 & 0.141 & 0.565 & 0.086 & 0.086 & 0.531 & 2.612 \\ 
&& & MC-SD & 0.334 & 0.149 & 0.567 & 0.090 & 0.091 & 0.465 & 2.089 \\
&& & CP (\%) & 94.0 & 91.3 & 94.0 & - & - & - & - \\
\cmidrule{2-11}
& \multirow{4}{*}{1.60} &  \multirow{4}{*}{5.10\%} & MC-Mean & 4.984 & 0.502 & 0.942 & -0.409 & 0.092 & 1.781 & 4.345 \\
&&& IM-SE & 0.319 & 0.141 & 0.557 & 0.091 & 0.088 & 0.533 & 3.326 \\ 
&&& MC-SD & 0.339 & 0.149 & 0.576 & 0.096 & 0.094 & 0.494 & 2.692 \\
&&& CP (\%) & 94.3 & 92.0 & 94.3 & - & - & - & - \\
\cmidrule{2-11}
& \multirow{4}{*}{3.45} &  \multirow{4}{*}{19.59\%} & MC-Mean & 4.977 & 0.505 & 0.931 & -0.418 & 0.089 & 1.812 & 4.559 \\
&&& IM-SE & 0.337 & 0.150 & 0.586 & 0.105 & 0.099 & 0.584 & 3.957 \\
&&& MC-SD & 0.349 & 0.156 & 0.588 & 0.110 & 0.103 & 0.516 & 2.914 \\
&&& CP (\%) & 93.3 & 93.7 & 93.3 & - & - & - & - \\
\cmidrule{2-11}
&  \multirow{4}{*}{4.30} & \multirow{4}{*}{34.11\%} & MC-Mean & 4.939 & 0.511 & 0.928 & -0.434 & 0.077 & 1.856 & 4.930 \\
&&& IM-SE & 0.371 & 0.163 & 0.644 & 0.122 & 0.115 & 0.721 & 4.872 \\
&&& MC-SD & 0.372 & 0.171 & 0.642 & 0.126 & 0.120 & 0.571 & 3.517 \\
&&& CP (\%) & 94.0 & 93.7 & 93.3 & - & - & - & - \\
\midrule
\multirow{16}{*}{300} &  \multirow{4}{*}{No} &  \multirow{4}{*}{0\%} & MC-Mean & 5.015 & 0.501 & 0.893 & -0.401 & 0.115 & 1.937 & 4.162 \\
&&& IM-SE & 0.167 & 0.084 & 0.299 & 0.048 & 0.048 & 0.293 & 1.172 \\ 
&&& MC-SD & 0.180 & 0.090 & 0.312 & 0.052 & 0.050 & 0.295 & 1.266 \\
&&& CP (\%)& 93.3 & 93.7 & 93.7 & - & - & - & - \\
\cmidrule{2-11}
& \multirow{4}{*}{1.60} & \multirow{4}{*}{5.31\%} & MC-Mean & 5.017 & 0.501 & 0.889 & -0.399 & 0.115 & 1.944 & 4.244 \\
&&& IM-SE & 0.169 & 0.085 & 0.303 & 0.053 & 0.051 & 0.314 & 1.444 \\
&&& MC-SD & 0.182 & 0.091 & 0.319 & 0.055 & 0.051 & 0.311 & 1.498 \\
&&& CP (\%)& 93.0 & 94.0 & 93.0 & - & - & - & - \\
\cmidrule{2-11} 
& \multirow{4}{*}{3.45} & \multirow{4}{*}{20.58\%} & MC-Mean & 4.999 & 0.507 & 0.891 & -0.409 & 0.110 & 1.997 & 4.645 \\
&&& IM-SE & 0.177 & 0.090 & 0.317 & 0.060 & 0.057 & 0.365 & 2.122 \\ 
&&& MC-SD & 0.191 & 0.095 & 0.332 & 0.061 & 0.058 & 0.341 & 1.935 \\
&&& CP (\%) & 93.7 & 92.3 & 93.7 & - & - & - & - \\
\cmidrule{2-11}
& \multirow{4}{*}{4.30}& \multirow{4}{*}{35.44\%} & MC-Mean & 4.958 & 0.518 & 0.901 & -0.409 & 0.110 & 2.080 & 5.043 \\
&&& IM-SE & 0.194 & 0.098 & 0.341 & 0.068 & 0.065 & 0.421 & 2.716 \\ 
&&& MC-SD & 0.199 & 0.106 & 0.339 & 0.074 & 0.069 & 0.385 & 2.258 \\
&&& CP (\%)& 95.0 & 93.3 & 95.3 & - & - & - & - \\
\midrule
\multirow{16}{*}{600} & \multirow{4}{*}{No} & \multirow{4}{*}{0\%} & MC-Mean & 5.001 & 0.507 & 0.909 & -0.400 & 0.118 & 1.966 & 4.072 \\ 
&&& IM-SE & 0.125 & 0.063 & 0.225 & 0.037 & 0.036 & 0.221 & 0.889 \\ 
&&& MC-SD & 0.125 & 0.064 & 0.226 & 0.038 & 0.037 & 0.212 & 0.918 \\
&&& CP (\%)& 93.0 & 94.7 & 92.3 & - & - & - & -  \\
\cmidrule{2-11}
& \multirow{4}{*}{1.60} &  \multirow{4}{*}{5.11\%} & MC-Mean & 4.999 & 0.508 & 0.911 & -0.401 & 0.117 & 1.977 & 4.145  \\
&&& IM-SE & 0.125 & 0.063 & 0.225 & 0.037 & 0.036 & 0.221 & 0.899 \\ 
&&& MC-SD & 0.124 & 0.065 & 0.224 & 0.038 & 0.037 & 0.212 & 0.937 \\
&&& CP (\%)& 92.7 & 94.3 & 92.0 & - & - & - & - \\
\cmidrule{2-11}
& \multirow{4}{*}{3.45} & \multirow{4}{*}{19.99\%} & MC-Mean & 4.981 & 0.513 & 0.921 & -0.405 & 0.117 & 2.034 & 4.447 \\
&&& IM-SE & 0.132 & 0.066 & 0.235 & 0.042 & 0.040 & 0.254 & 1.252 \\
&&& MC-SD & 0.127 & 0.069 & 0.229 & 0.042 & 0.041 & 0.239 & 1.342 \\
&&& CP (\%)& 94.0 & 93.7 & 94.3 & - & - & - & - \\
\cmidrule{2-11}
& \multirow{4}{*}{4.30} & \multirow{4}{*}{34.64\%} & MC-Mean & 4.939 & 0.521 & 0.933 & -0.410 & 0.112 & 2.113 & 4.826 \\
&&& IM-SE & 0.142 & 0.072 & 0.251 & 0.048 & 0.045 & 0.300 & 1.672 \\
&&& MC-SD & 0.139 & 0.074 & 0.251 & 0.049 & 0.046 & 0.284 & 1.657 \\
&&& CP (\%)& 93.7 & 94.7 & 94.7 & - & - & - & - \\
\bottomrule
\end{tabular*}
\end{table}

\subsection{Simulation study 1}
This study aims to provide empirical evidence about the consistence of the ML estimates under different scenarios. Thence, 300 MC samples were generated considering different sample sizes: $n=100, 300$, and $600$. The data was generated from the model specified by Equations \ref{modelo1}-\ref{error1}, considering a Student-$t$ distribution for the innovations ($\eta$'s) with $\sigma^2 = 2$ and $\nu=4$. The rest of the parameters were set as $\bbeta=(5, 0.50, 0.90)^\top$, $\bphi=(-0.40, 0.12)^\top$, and the covariables at time $i$ were $\x_i = (1,x_{i1},x_{i2})^\top$, with $x_{i1}\sim\N(0,1)$ and $x_{2i}\sim\mathcal{U}(0, 1)$, for $i=1,2,\ldots,n$. From this scenario, two analysis were conducted and will be discussed next.

\subsubsection{Asymptotic properties}

Aiming to have scenarios with an average level of censoring/missing of 5\%, 20\%, and 35\%, respectively, we considered the following LODs (values lower than the LOD are substituted by the LOD): 1.60, 3.45, and 4.30. Furthermore, 20\% of the desired censored rate corresponds to observations randomly selected to be treated as missing. Additionally, we considered the case without censoring (original data) for comparison.

For each sample size and LOD, we computed the mean of the 300 MC estimates  (MC-Mean), the mean of the standard error (\textrm{IM-SE}) computed by the inverse of the observed information matrix given in Equation \ref{IMSE}, the standard deviation of the 300 MC estimates (MC-SD), and the coverage probability (\textrm{CP}) of a 95\% confidence interval, i.e.,
\begin{small}\begin{eqnarray*}
\mbox{MC-Mean}_i=\bar{\widehat{\theta}}_i = \frac{1}{300}\sum_{j=1}^{300} \widehat{\theta}_i^{(j)}, \quad \mbox{MC-SD}_i=\sqrt{\frac{1}{299}\sum_{j=1}^{299}\left(\widehat{\theta}_i^{(j)}-\bar{\widehat{\theta}}_i\right)^2}, \quad \mbox{and} \quad \mbox{IM-SE}_i=\frac{1}{300}\sum_{j=1}^{300} SE\left(\widehat{\theta}_i^{(j)}\right),
\end{eqnarray*}\end{small}
\noindent where $\widehat{\theta}_i^{(j)}$ is the estimate of the $i$th parameter of $\btheta = (\beta_0, \beta_1, \beta_2, \phi_1, \phi_2, \sigma^2, \nu)^\top$ in the $j$th MC sample.

\begin{figure}[h]
\caption{\textbf{Simulation study 1}. Boxplot of the estimates for CAR$t$(2) model considering different sample sizes and LOD.} \label{boxplot1}
\centering
\includegraphics[scale=.7]{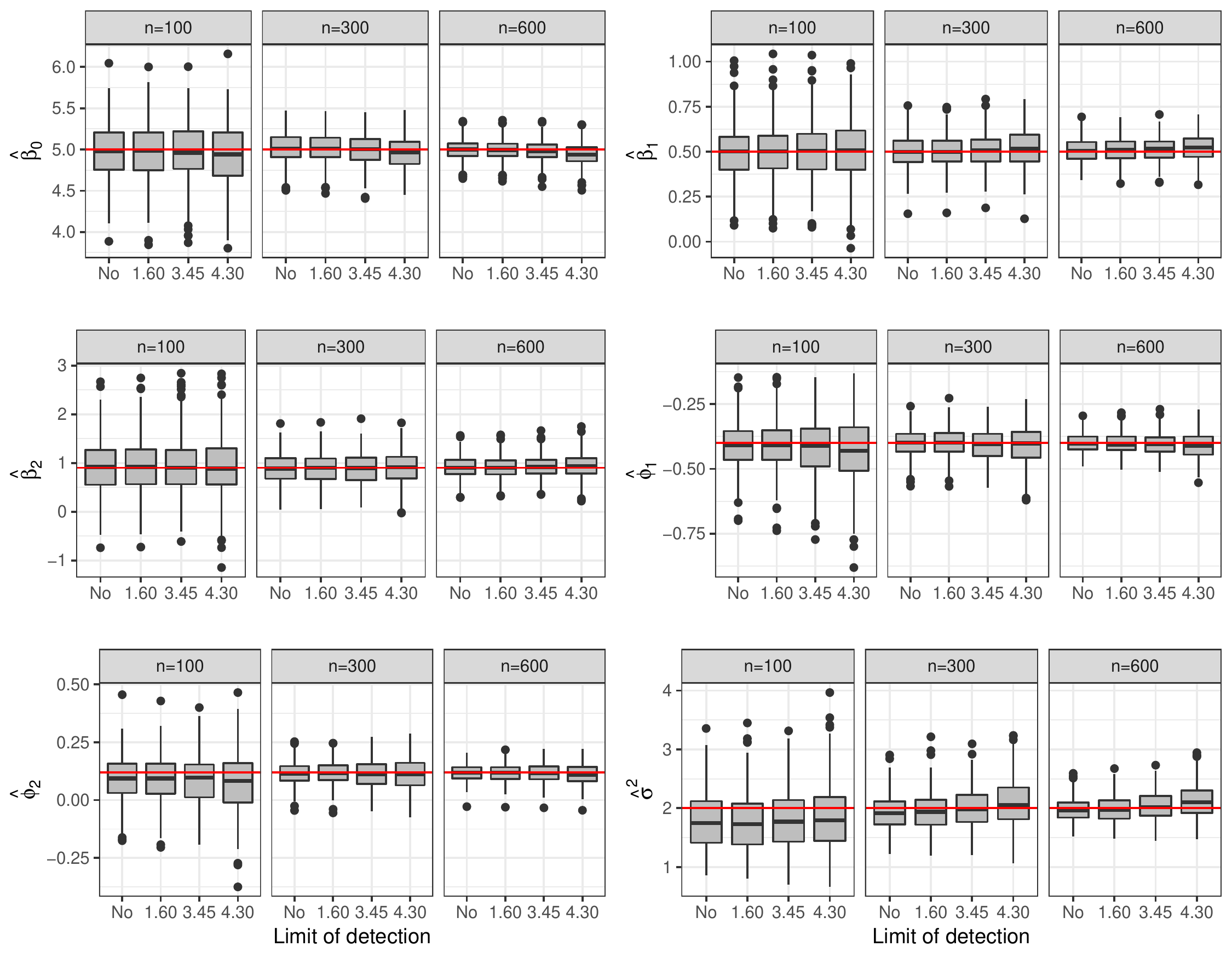}
\end{figure}

The results are shown in Table \ref{estimates1}, where we can observe that the mean of the estimates (\textrm{MC-mean}) is close to the true parameter value in all combinations of sample size and LOD, except for the scale parameter $\sigma^2$ for a sample size of  $n=100$. As expected, this difference decreases as the sample size increases. Notice that the mean of the standard errors obtained through the inverse of the observed information matrix (\textrm{IM-SE}) are in general close to the standard deviation of the estimates (\textrm{MC-SD}) for all scenarios, indicating that the proposed method to obtain the standard errors is reliable.

Figure \ref{boxplot1} shows the boxplots of the estimates for each parameter, considering different sample sizes and LODs. The solid red line represents the true parameter value. In general, the median of the estimates is close to the real value independent of the sample size and LOD. However, for $\phi_2$ and $\sigma^2$, the median underestimates the true value in samples of size $n = 100$, i.e., the smallest sample size in the simulation study. Furthermore, interquartile ranges decrease as sample sizes increase, suggesting the consistence of the estimates. Additionally, boxplots for the estimates of $\nu$ are shown in Figure \ref{est1:nu} in Appendix \ref{sec:ap:fig}.

\begin{figure}[h]
\centering
\caption{\textbf{Simulation study 1}. MSE of the estimates for the CAR$t$(2) model considering different sample sizes and LOD.} \label{MSE1}
\includegraphics[scale=.75]{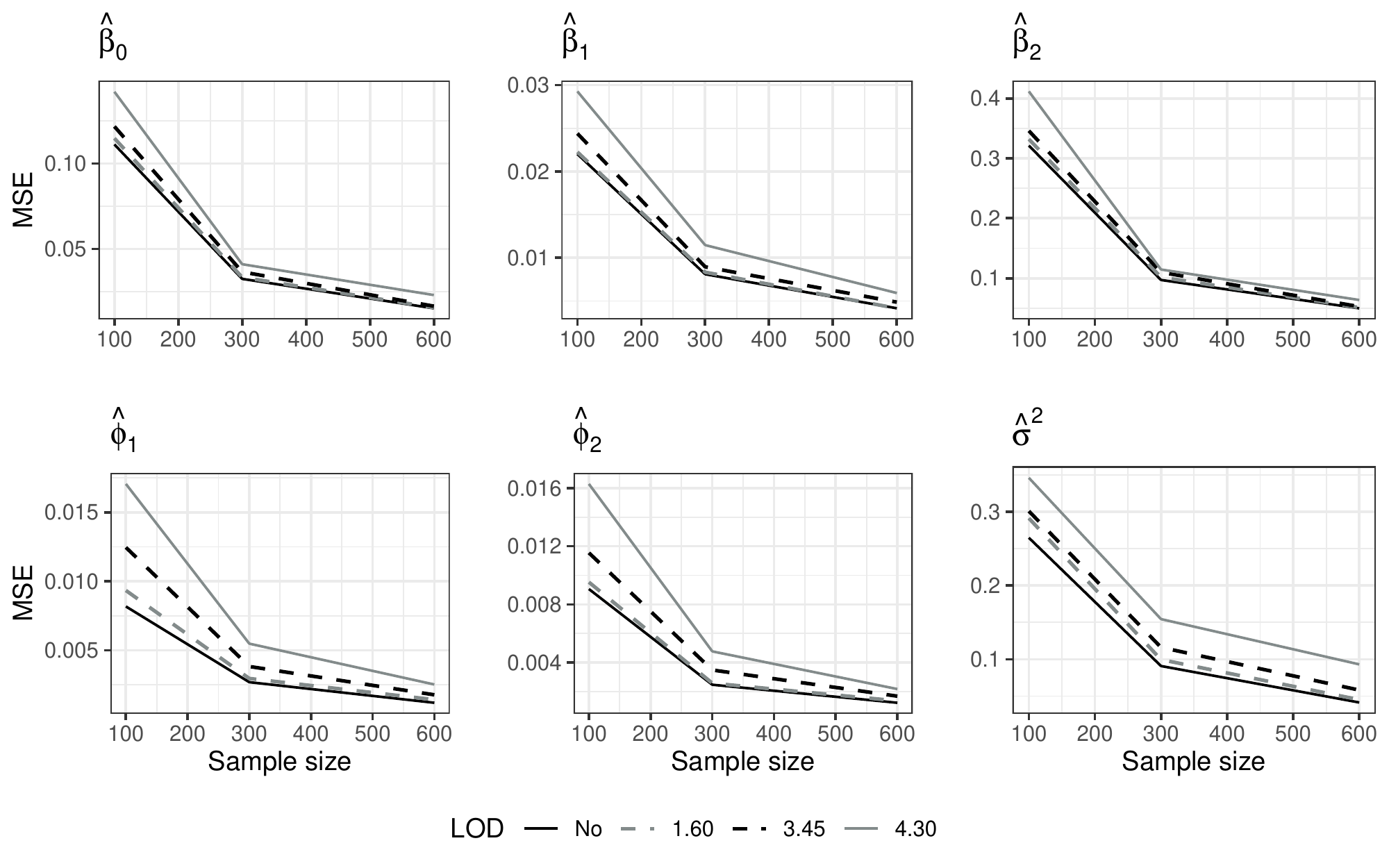}
\end{figure}

Finally, to study the asymptotic properties of the estimates, we analyzed the mean squared error (MSE) of the estimates obtained from the proposed algorithm for all scenarios, which can defined by $\mbox{MSE}_i = \frac{1}{n}\sum_{j=1}^{n} (\widehat{\theta}_i^{(j)} - \theta_i )^2$. The results for each parameter, sample size, and LOD are shown in Figure \ref{MSE1}, where we may note that the MSE tends to zero as the sample size increases. Thus, the proposed SAEM algorithm seems to provide ML estimates with good asymptotic properties for our proposed autoregressive censored linear model with Student-$t$ innovations.

%%%%%%%%%%%%%%%%%%%%%%%%%%%%%%%%%%%%%%% 

\subsubsection{Residual Analysis}\label{residual}

Checking the specification of a statistical model usually involves statistical tests and graphical methods based on residuals. However, conventional residuals (as Pearson's residuals, p.e.) are not appropriate for some models, since they may lead to erroneous inference as \cite{kalliovirta2006misspecification} demonstrated for models based on mixtures of distributions. As an alternative, \cite{dunn1996randomized} developed the quantile residuals, a type of residuals for regression models with independent responses that produce residuals normally distributed by inverting the fitted distribution function for each response value and finding the equivalent standard normal quantile. These results assume that the model is correctly specified and parameters are consistently estimated. The method can be extended to dependent data by expressing the likelihood as a product of univariate conditional likelihoods.

To compute the quantile residuals for the CAR$t(p)$ model, we first impute the censored or missing observations as defined in Section \ref{sec:prediction}. Then, considering all values as completely observed, the residual for $i$th observation is computed as
follows
\begin{equation}\label{eq_resid}
\widehat{r}_i = \Phi^{-1}\left( T_1\left( y_i ; \widehat{\mu}_i, \widehat{\sigma}^2, \widehat{\nu} \right) \right), \quad i=p+1,\ldots n,
\end{equation}
where $\Phi(\cdot)$ denotes the cumulative distribution function (CDF) of the standard normal distribution, $T_1(\cdot; \hat{\mu}_i, \hat{\sigma}^2, \hat{\nu})$ is the CDF of the univariate Student-$t$ distribution with location parameter $\widehat{\mu}_i = \x_i^\top \widehat{\bbeta} + (\y_{(i,p)} - \X_{(i,p)} \widehat{\bbeta})^\top \widehat{\bphi}$, scale parameter $\hat{\sigma}^2$, and $\hat{\nu}$ degrees of freedom, where $\widehat{\btheta}$ refers to the ML estimates of $\btheta$ obtained through the SAEM algorithm. Note that the quantile residual is calculated only for the latest $n-p$ observations.

\begin{figure}[ht]
\centering
\caption{\textbf{Simulation study 1}. Plots of quantile residuals for a sample of size $n=600$ generated from the CAR$t$(2) model considering different levels of left censoring.} \label{residualSim}
\includegraphics[scale=.68]{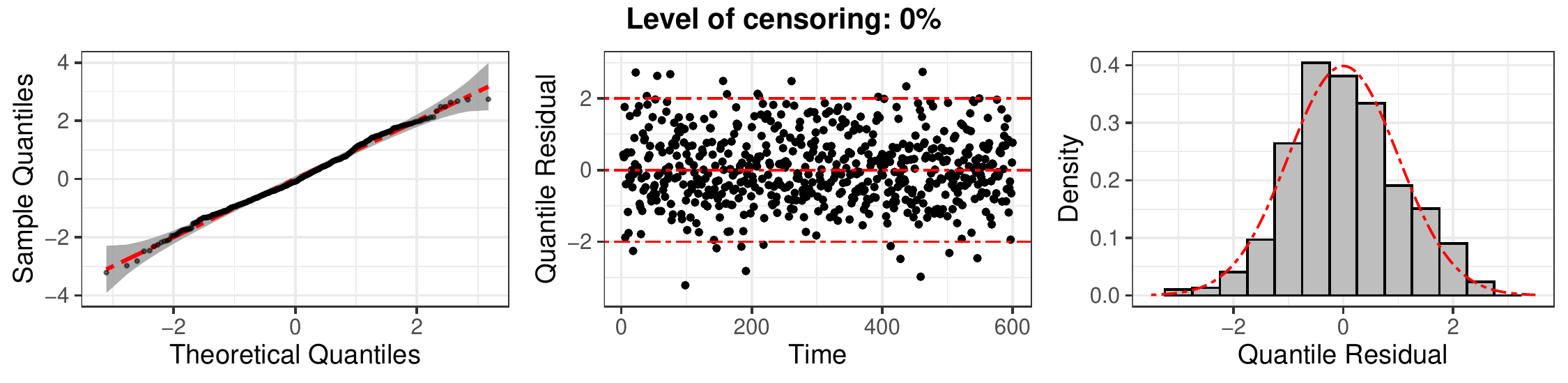}
\includegraphics[scale=.68]{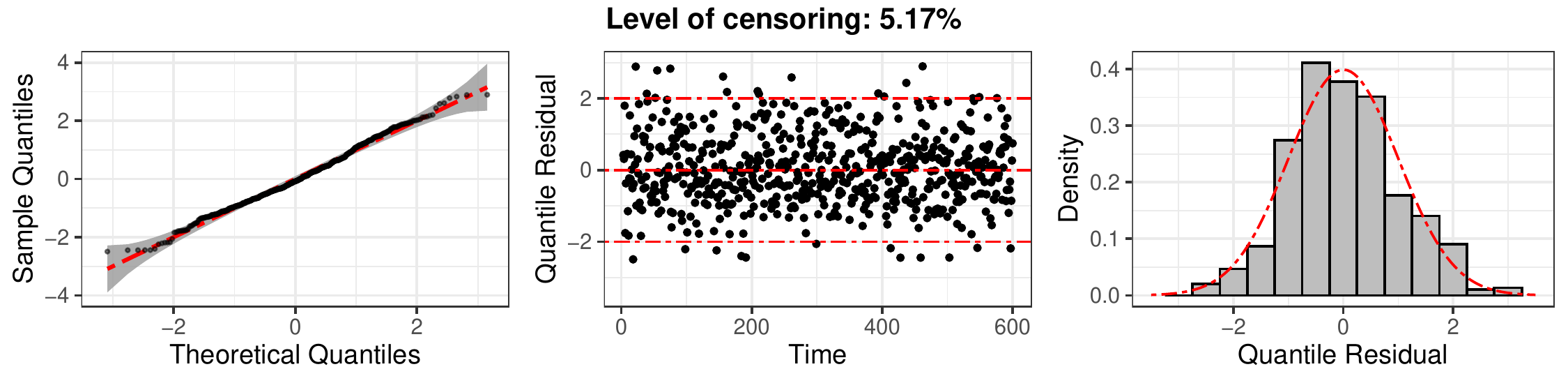}
\includegraphics[scale=.68]{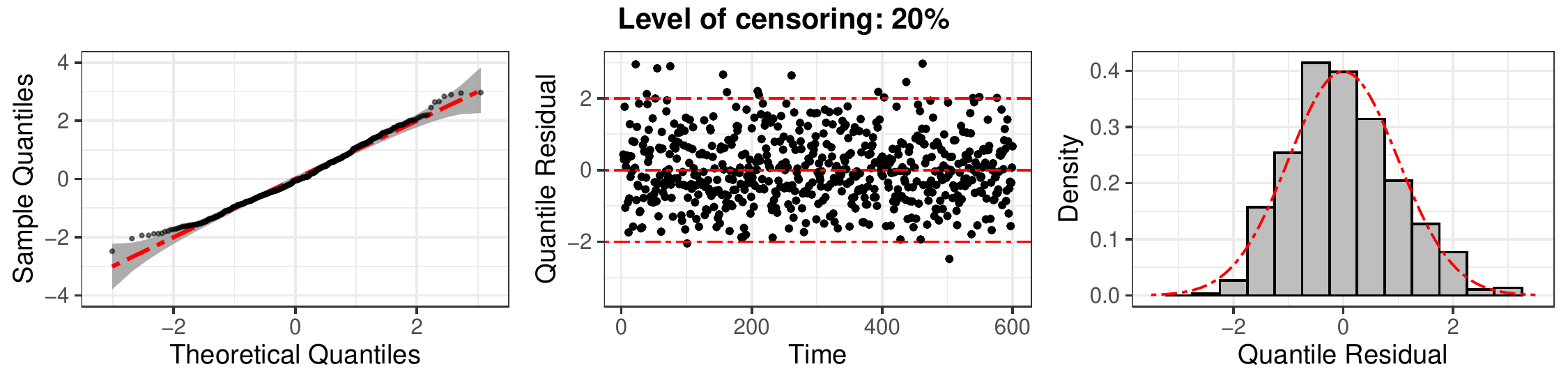}
\includegraphics[scale=.68]{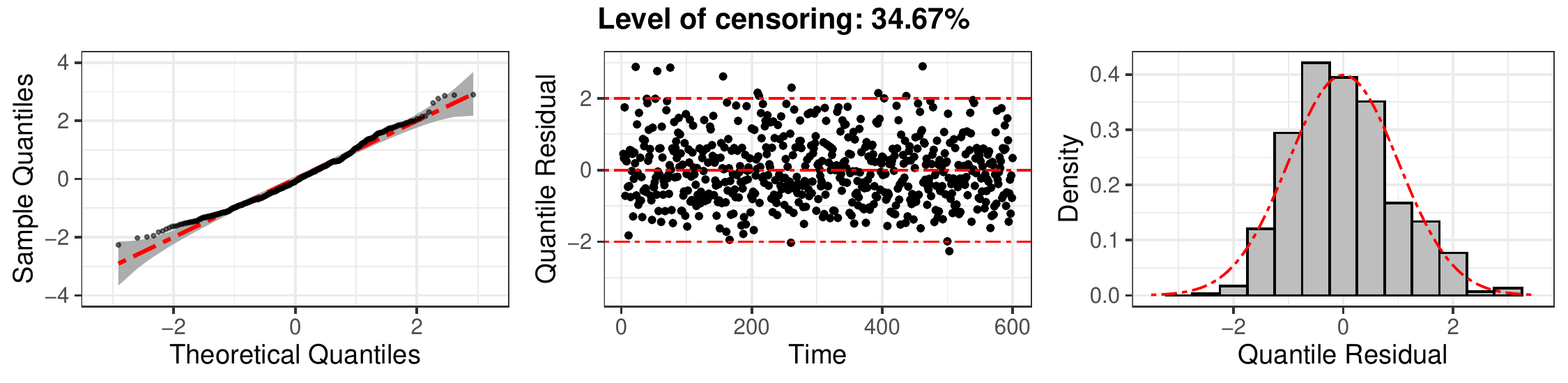}
\end{figure}

Aiming to analyze how the quantile residuals behave for the proposed model, they were computed for a simulated dataset of sample size $n=600$ and considering four levels of left censoring: 0\%, 5.17\%, 20\%, and 34.67\%. Figure \ref{residualSim} shows a Quantile-Quantile (Q-Q) plot, a dispersion plot (residual vs. time), and a histogram for the residuals. For all levels of censoring, the histogram seems to correspond to a histogram of a normally distributed variable, and the dispersion plot shows independent residuals. We can deduce through the Q-Q plot that the residuals are roughly normally distributed because all points form a roughly straight line inside the confidence band. However, for samples with 20\% and 34.67\% of censoring, the Q-Q plots present a slight deviation from the center line in the lower tail, which might be due to the high proportion of censored values.

\begin{figure}[h]
\centering
\caption{\textbf{Simulation study 1}. Plots of the residuals for a sample of size $n=600$ generated from the CAR$t$(2) model and fitting a model with normal innovations, considering different levels of left censoring.} \label{residualNorm}
\includegraphics[scale=.68]{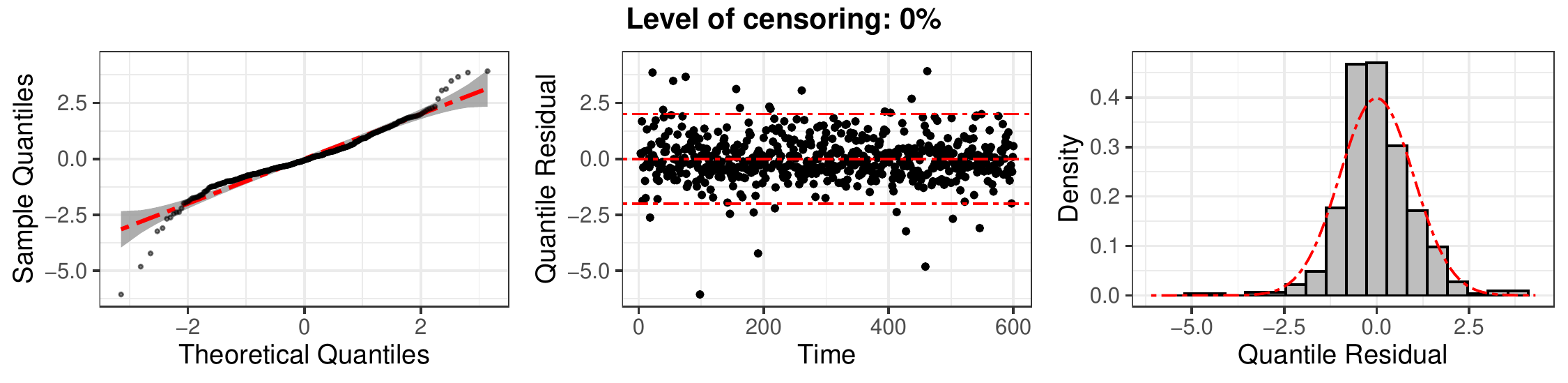}
\includegraphics[scale=.68]{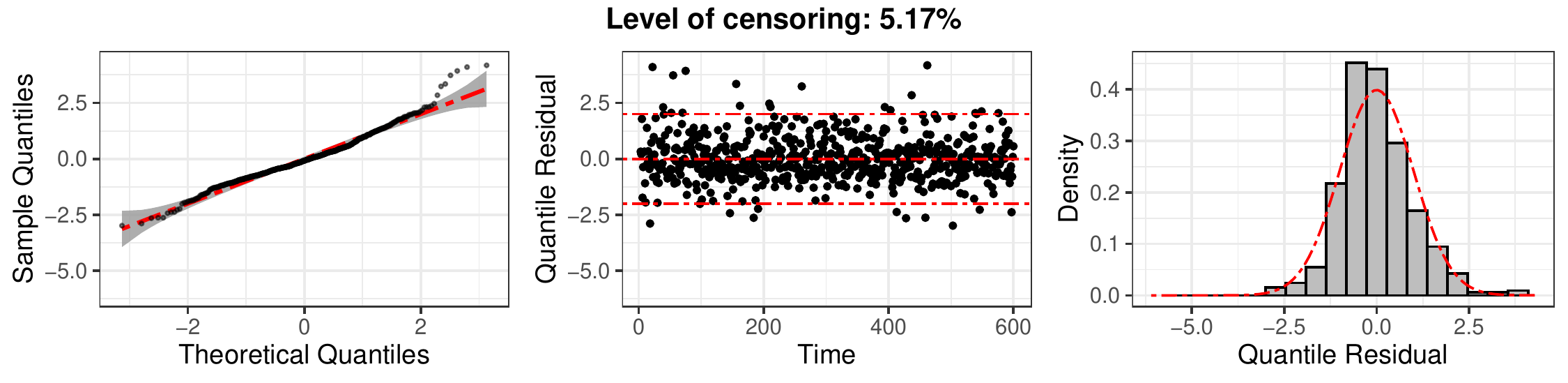}
\includegraphics[scale=.68]{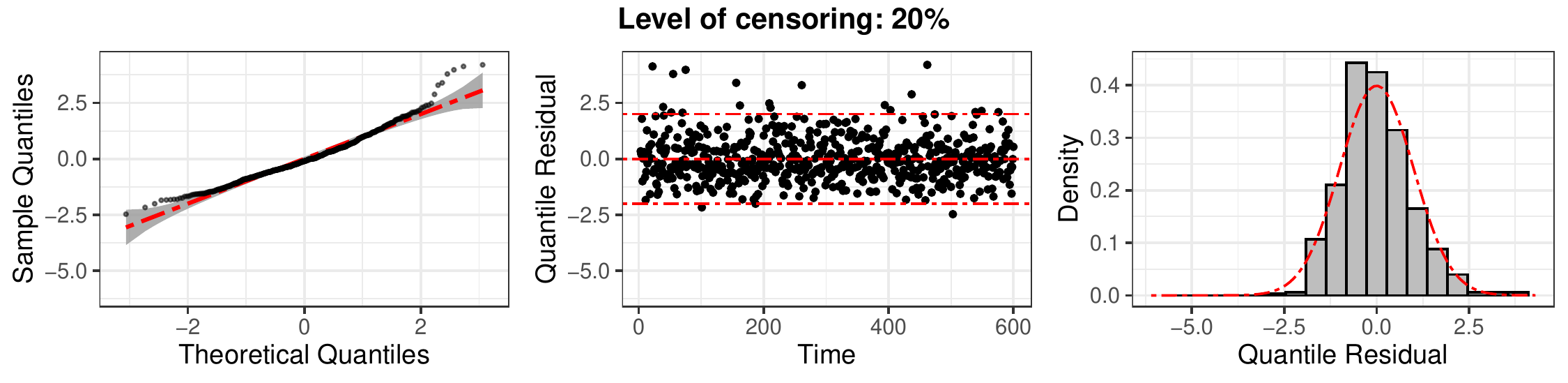}
\includegraphics[scale=.68]{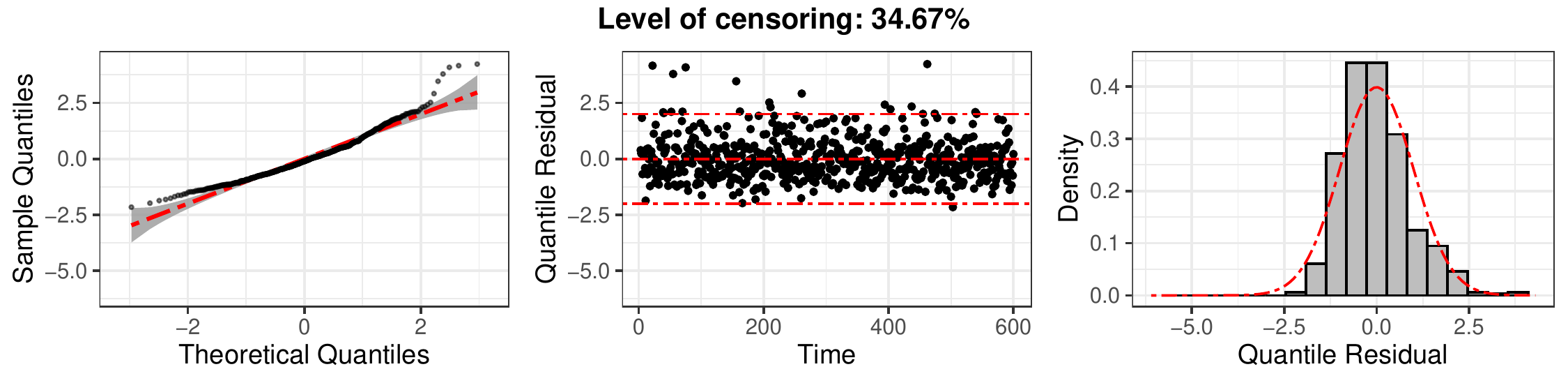}
\end{figure}

For comparison, we fitted the same dataset considering the normal distribution (i.e., disregarding the heavy tails), and computed the corresponding quantile residuals. The resulting plots are given in Figure \ref{residualNorm}, where we can see clear signs of non-normality, such as large residuals and several points outside the confidence band in the Q-Q plots. Therefore, this illustration indicates that this method can help checking the model specification in the CAR$t(p)$ model. Nevertheless, for significant levels of censoring, more caution is needed in the analysis of residuals since our proposal imputes the unobserved data by its conditional expectation.

%%%%%%%%%%%%%%%%%%%%%%%%%%%%%%%%%%%%%%%%%%

\subsection{Simulation study 2: Robustness of the estimators}

This simulation study aims to compare the performance of the estimates for two censored AR models in the presence of outliers on the response variable. In this case, we generated 300 MC samples of size $n=100$ under the model specified in Equations \ref{modelo1}-\ref{error1}, considering the standard normal distribution %, $\mathcal{N}(0,1)$, 
for the innovations. The other parameters were set as $\bbeta=(4, 0.50)^\top$ and $\bphi=(0.48, -0.20)^\top$, and the covariates were set as $\x_i=(1,x_{i1})^\top$, where $x_{i1}\sim\N(0,1)$, $i = 1, \ldots, 100$. After generating the data, each MC sample was perturbed considering the following scheme:  the maximum value was increased in $\vartheta$ times the sample standard deviation, i.e., $\max(\y) = \max(\y) + \vartheta \mbox{SD}(\y)$, for $\vartheta\in\{0,1,2,3,4,5,7\}$. 

Furthermore, we considered three different levels of censoring: the first case corresponds to the case without censoring; the second case considered 2.34 as a LOD, where all values lower or equal than the LOD were substituted by 2.34 (which implied an average of 10.04\% of censoring); and finally the third scenario considered a LOD of 3.31 (which yielded an average of 30\% of censoring). For each scenario, we fitted two models: the first considers normal innovations, denoted by CAR(2) whose parameters were estimated by the function \texttt{ARCensReg} from the \textsf{R} package \textbf{ARCensReg} \citep{schumacher2016package}, and the second one is our proposed model CAR$t$(2) fitted through the function \texttt{ARtCensReg} from the same package. 

Table \ref{perturbed1} displays the mean of the estimates for each parameter by level of perturbation and censoring rate. To obtain comparable values of $\sigma^2$, for the model with Student-$t$ innovations we reported the estimated variance of the innovation $\widehat{\sigma}^{2*} = \widehat{\nu}\widehat{\sigma}^{2}/(\widehat{\nu} - 2)$, where $\widehat{\sigma}^{2}$ is the estimate of the scale parameter under our proposal. Moreover, the percentage of times that our model detected the perturbed observation as influential, denoted by DI(\%), is also reported, which was computed as the number of times the estimated weight (${\widehat{u}_i}$, computed during the E-step in the SAEM algorithm) of the perturbed observation was the lowest one, divided by the number of MC samples. 

\begin{table}[ht]
\caption{\textbf{Simulation study 2}. Mean of the estimates for CAR(2) and CAR$t$(2) model based on 300 MC samples of size $n=100$ for different levels of perturbation.} \label{perturbed1}
{a. Level of censoring: 0\%}
\centering
\small
\begin{tabular*}{\textwidth}{c@{\extracolsep{\fill}}ccccccccccccc}
\toprule
Pert. & \multicolumn{5}{c}{CAR(2)} & & \multicolumn{7}{c}{CAR$t$(2)}\\
\cline{2-6} \cline{8-14}
($\vartheta$) & $\beta_0$ & $\beta_1$ & $\sigma^2$ & $\phi_1$ & $\phi_2$ & & $\beta_0$ & $\beta_1$ & $\sigma^{2*}$ & $\phi_1$ & $\phi_2$ & $\nu$ & DI (\%)\\ 
\midrule
0 & 4.007 & 0.510 & 0.963 & 0.478 & -0.215 & & 4.007 & 0.510 & 0.979 & 0.480 & -0.216 & 24.424 & 17 \\ 
1 & 4.021 & 0.526 & 1.030 & 0.469 & -0.209 & & 4.008 & 0.518 & 1.042 & 0.464 & -0.207 & 17.098 & 73.33 \\ 
2 & 4.036 & 0.541 & 1.135 & 0.447 & -0.195 & & 3.999 & 0.518 & 1.130 & 0.433 & -0.186 & 9.568 & 98.67\\ 
3 & 4.050 & 0.557 & 1.275 & 0.416 & -0.175 & & 3.991 & 0.515 & 1.229 & 0.398 & -0.162 & 6.064 & 99.67 \\ 
4 & 4.064 & 0.572 & 1.449 & 0.383 & -0.154 & & 3.987 & 0.513 & 1.328 & 0.369 & -0.140 & 4.896 & 100\\ 
5 & 4.079 & 0.588 & 1.655 & 0.349 & -0.133 & & 3.986 & 0.510 & 1.426 & 0.346 & -0.121 & 4.324 & 100\\ 
7 & 4.107 & 0.619 & 2.160 & 0.287 & -0.100 & & 3.985 & 0.508 & 1.628 & 0.309 & -0.090 & 3.710 & 100\\ 
\bottomrule
\end{tabular*}

\vspace*{.2cm}

{b. Level of censoring: 10.04\%} \label{perturbed2}
\centering\small
\begin{tabular*}{\textwidth}{c@{\extracolsep{\fill}}ccccccccccccc}
\toprule
Pert. & \multicolumn{5}{c}{CAR(2)} & & \multicolumn{7}{c}{CAR$t$(2)}\\
\cline{2-6} \cline{8-14}
($\vartheta$) & $\beta_0$ & $\beta_1$ & $\sigma^2$ & $\phi_1$ & $\phi_2$ & & $\beta_0$ & $\beta_1$ & $\sigma^{2*}$ & $\phi_1$ & $\phi_2$ & $\nu$ & DI (\%) \\
\midrule
0 & 4.006 & 0.511 & 0.966 & 0.478 & -0.214 & & 4.006 & 0.511 & 0.997 & 0.478 & -0.213 & 20.892 & 21.00 \\ 
1 & 4.018 & 0.529 & 1.044 & 0.468 & -0.208 & & 4.007 & 0.519 & 1.079 & 0.460 & -0.203 & 14.290 & 77.67 \\ 
2 & 4.027 & 0.549 & 1.165 & 0.445 & -0.194 & & 4.002 & 0.518 & 1.207 & 0.424 & -0.180 & 7.764 & 98.67 \\ 
3 & 4.036 & 0.570 & 1.327 & 0.415 & -0.175 & & 3.999 & 0.514 & 1.362 & 0.387 & -0.155 & 5.061 & 99.67 \\ 
4 & 4.044 & 0.592 & 1.527 & 0.383 & -0.156 & & 3.999 & 0.511 & 1.535 & 0.357 & -0.133 & 4.105 & 100 \\ 
5 & 4.052 & 0.614 & 1.764 & 0.351 & -0.138 & & 4.000 & 0.509 & 1.704 & 0.332 & -0.115 & 3.668 & 100 \\ 
7 & 4.065 & 0.659 & 2.340 & 0.294 & -0.109 & & 4.003 & 0.505 & 2.170 & 0.295 & -0.088 & 3.166 & 100 \\ 
\bottomrule
\end{tabular*}

\vspace*{.2cm}

{c. Level of censoring: 30\%}\label{perturbed3}
\centering\small
\begin{tabular*}{\textwidth}{c@{\extracolsep{\fill}}ccccccccccccc}
\toprule
Pert. & \multicolumn{5}{c}{CAR(2)} & & \multicolumn{7}{c}{CAR$t$(2)}\\
\cline{2-6} \cline{8-14}
($\vartheta$) & $\beta_0$ & $\beta_1$ & $\sigma^{2}$ & $\phi_1$ & $\phi_2$ & & $\beta_0$ & $\beta_1$ & $\sigma^{2*}$ & $\phi_1$ & $\phi_2$ & $\nu$ & DI (\%) \\ \midrule
0 & 4.014 & 0.505 & 0.944 & 0.476 & -0.219 & & 4.015 & 0.503 & 0.972 & 0.476 & -0.218 & 19.338 & 24.67 \\ 
1 & 4.014 & 0.532 & 1.050 & 0.464 & -0.213 & & 4.011 & 0.513 & 1.076 & 0.454 & -0.206 & 12.553 & 82.33 \\ 
2 & 4.007 & 0.562 & 1.213 & 0.439 & -0.199 & & 4.007 & 0.512 & 1.230 & 0.415 & -0.181 & 6.854 & 98.33 \\ 
3 & 3.996 & 0.594 & 1.429 & 0.409 & -0.182 & & 4.006 & 0.507 & 1.406 & 0.376 & -0.156 & 4.651 & 100 \\ 
4 & 3.982 & 0.628 & 1.694 & 0.378 & -0.165 & & 4.008 & 0.503 & 1.616 & 0.344 & -0.135 & 3.803 & 100 \\ 
5 & 3.965 & 0.664 & 2.005 & 0.349 & -0.151 & & 4.010 & 0.500 & 1.837 & 0.318 & -0.118 & 3.386 & 100 \\ 
7 & 3.926 & 0.737 & 2.760 & 0.301 & -0.130 & & 4.016 & 0.495 & 2.428 & 0.280 & -0.092 & 2.926 & 100 \\ 
\bottomrule
\end{tabular*}
\end{table}

\begin{figure}[ht]
\caption{\textbf{Simulation study 2}. Boxplot of the estimates obtained from CAR(2) and CAR$t$(2) model based on 300 MC samples of size 100.} \label{box_pert}
\centering
{a. Level of censoring: 0\%}
\includegraphics[scale=.66]{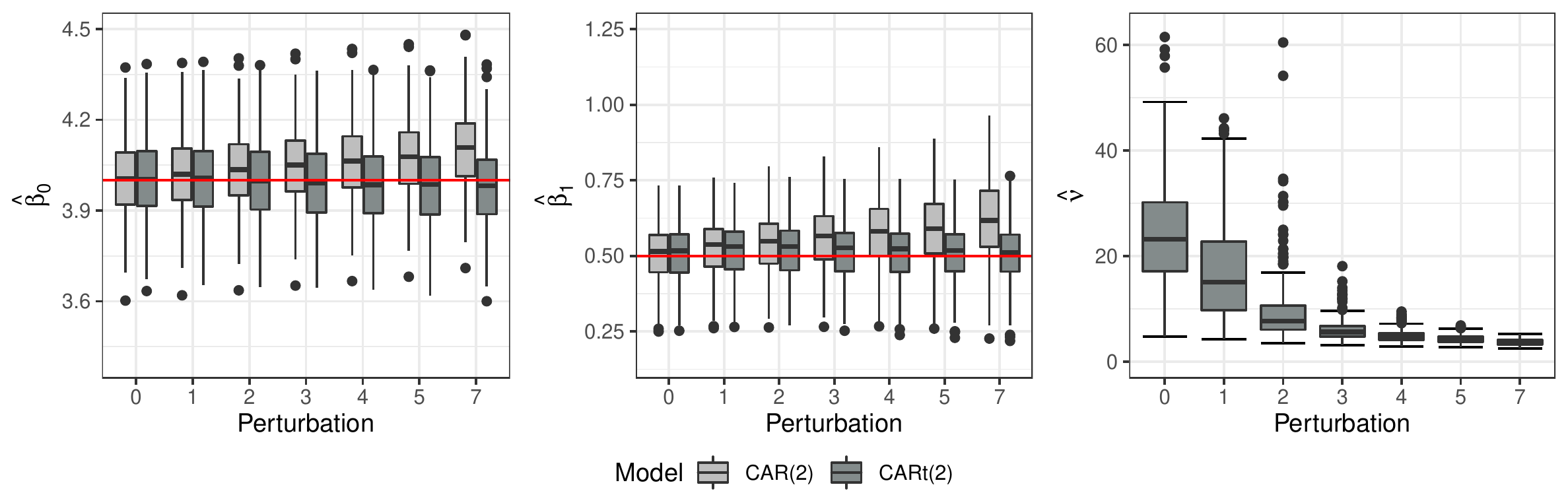}

{b. Level of censoring: 10.04\%}
\includegraphics[scale=.66]{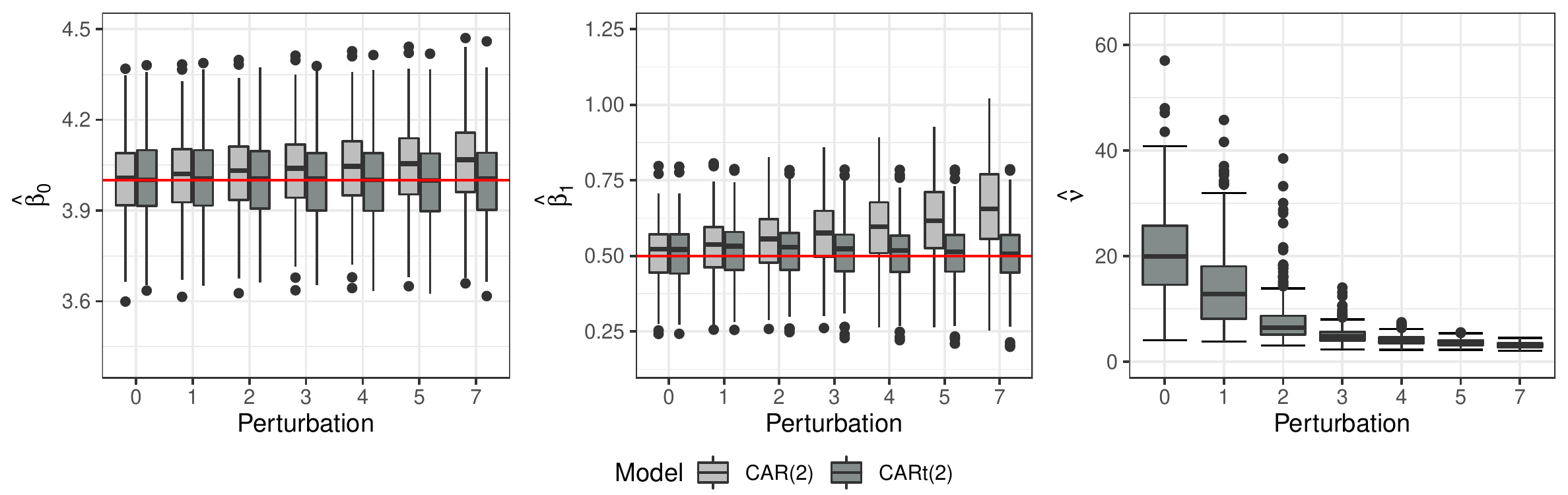}

{c. Level of censoring: 30\%}
\includegraphics[scale=.66]{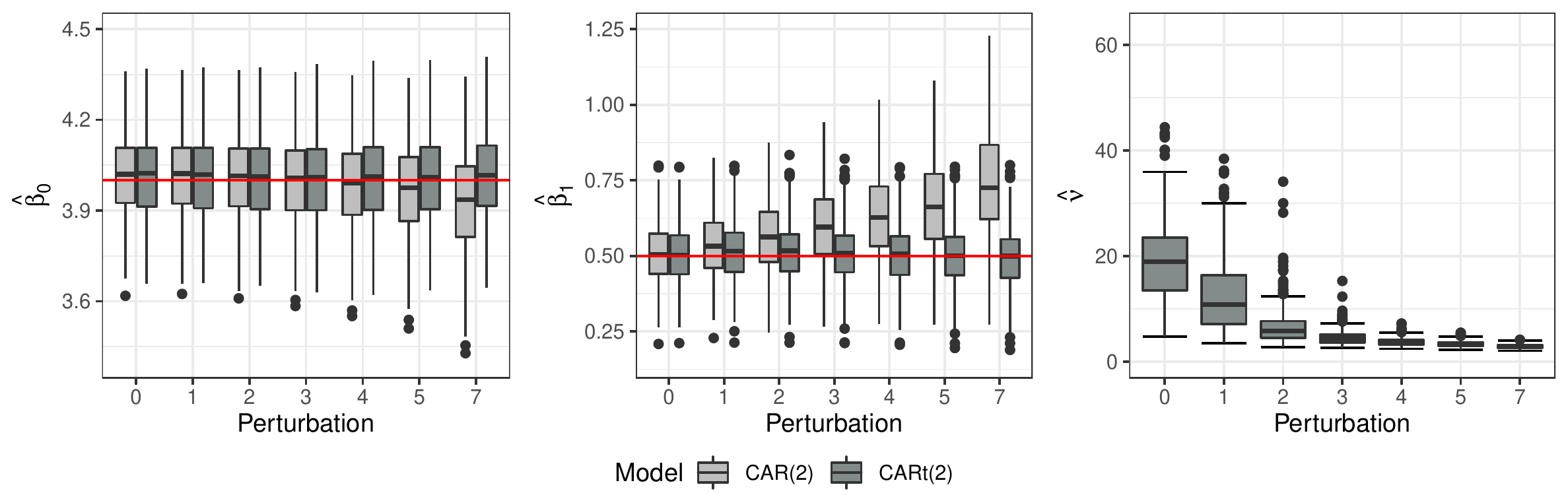}
\end{figure}

Table \ref{perturbed1}a and Figure \ref{box_pert}a show the results for the non-censored dataset, where it is possible to observe that for the normal distribution the bias increases as the perturbation increases, as natural; however, when Student-$t$ innovations are considered, the bias is much smaller, illustrating the robustness of the model against atypical distributions. As expected, estimates for $\nu$ decrease as the perturbation grows. For the non-perturbed samples, the observation with the maximum value was detected as influential in only 17\% of the samples, but this percentage increases fast along the perturbation. Observe that for $\vartheta = 4, 5$, and $7$, the perturbed observation was detected as influential in all MC samples.

Table \ref{perturbed1}b and Figure \ref{box_pert}b display the results for samples with an average of 10.04\% of censoring. The estimates for $\beta_0$ have a distribution similar to the non-censored case. On the other hand, for $\beta_1$, a more significant difference  was observed between the real value and its estimate in the normal model. In contrast, the model with heavy-tailed innovations performs better in recovering the true parameter values, with a mean value of $\nu$ smaller than the observed previously.

Finally, results for the scenario with an average of 30\% of left censoring are shown in Table \ref{perturbed1}c and Figure \ref{box_pert}c. For the normal case, the bias for $\beta_1$ is more outstanding than the observed in the previous two cases, while for our model, its mean and median are generally close to the true parameter value, for all perturbation levels. For $\beta_0$, the normal model returned estimates close to the real value only for levels of perturbation lower than 4 ($\vartheta < 4$), while for larger perturbations the models tends to underestimate it. These results confirm that the heavy-tails of the Student-$t$ distribution provides to our model the ability to mitigate the effect of outliers, i.e., a much more robust method against atypical values. %In Figure \ref{box_pert} this last can be confirmed by studying the estimate of $\nu$ under the three scenarios where $\nu$ decrease (the tails of the Student-$t$ distribution become thicker) as the perturbation increases.

%%%%%%%%%%%%%%%%%%%%%%%%%%%%%%%%%%%%%%

\section{Application}\label{application}

In this section, the CAR$t(p)$ model is applied to a real environmental dataset that has both missing and censored observations. We consider the phosphorus concentration dataset previously analyzed by \cite{schumacher2017censored} and \cite{wang2018quasi}, available in the package \textbf{ARCensReg}. For this study, we are interested in tracking the phosphorus concentration levels over time as an indicator of the river water quality since, for instance, excessive phosphorus in surface water may result in eutrophication. The data consist of $n=181$ observations of phosphorus concentration ($P$) in mg/L monthly measured from October 1998 to October 2013 in West Fork Cedar River at Finchford, Iowa, USA. The phosphorus concentration measurement was subject to a LOD of 0.02, 0.05, or 0.10, depending on the time, and therefore 28 (15.47\%) observations are left-censored. Moreover, there are 7 (3.87\%) missing observations from September 2008 to March 2009 due to a program suspension caused by a temporary lack of funding. 

As mentioned by \cite{wang2018quasi}, $P$ levels are generally correlated with the water discharge ($Q$), which is measured in cubic feet per second; then, our objective is to explore the relationship between $P$ and $Q$, when the response contains censored and missing observations. Aiming to evaluate the prediction accuracy, the dataset was train-test split. The training dataset consists of 169 observations, where 20.71\% are left-censored or missing, while the testing dataset contains 12 observations, all fully observed. 
Following \cite{wang2018quasi}, to make the $P$-$Q$ relationship more linear we considered the logarithmic transformation of $P$ and $Q$ and fitted the following CAR$t(p)$ model:
\begin{equation}
\label{app2}
\log(P_t) = \sum_{j=1}^4 \left[ \beta_{0j}S_{jt} + \beta_{1j}S_{jt}\log(Q_t) \right] + \xi_t, \quad t=p+1,\ldots,n,
\end{equation}
where the regression error $\xi_t$ follows an autoregressive model and $S_{j}$ is a dummy seasonal variable for quarters $j=1,2,3$, and $4$. The first quarter corresponds from January to March, the second one from April to June, and so on. In this model, $\beta_{0j}$ and $\beta_{1j}$ are respectively the intercept and slope for quarter $j$, for $j=1,2,3$, and $4$. 

The AR order is unknown, but since the second observation from the training set (November 1998) is censored, it is not possible to consider a model of order greater than one because the proposed algorithm assumes that the first $p$ values are completely observed. Therefore, for this application we only considered a model of order 1. Besides that \cite{schumacher2017censored} concluded that a censored autoregressive model of order 1 was the best to fit this dataset based on information criteria and mean squared prediction error (MSPE). The authors also found that the dataset has some influential observations, then it seems reasonable to consider a model with innovations following a heavy-tailed distribution.

\begin{table}[ht]
\centering
\caption{\textbf{Phosphorus concentration data}. Parameter estimates and their standard errors (SE) for the CAR(1) and CAR$t$(1) models.} \label{app2_est}
\small
\begin{tabular*}{\textwidth}{l@{\extracolsep{\fill}}cccccc}
\toprule
\multirow{2}{*}{Parameters} & \multicolumn{3}{c}{CAR$t$(1)} & \multicolumn{3}{c}{CAR(1)} \\
\cmidrule{2-4} \cmidrule{5-7}
& Estimate & SE & 95\% CI & Estimate & SE & 95\% CI \\
\midrule
$\beta_{01}$ & -4.338 & 0.609 & (-5.530 , -3.145) & -4.670 & 0.513 & (-5.674 , -3.665) \\
$\beta_{02}$ & -2.731 & 0.750 & (-4.201 , -1.262) & -3.022 & 0.751 & (-4.494 , -1.549) \\
$\beta_{03}$ & -4.141 & 0.419 & (-4.963 , -3.319) & -4.174 & 0.442 & (-5.041 , -3.307) \\
$\beta_{04}$ & -4.651 & 0.545 & (-5.719 , -3.583) & -4.999 & 0.549 & (-6.075 , -3.922) \\
$\beta_{11}$ &  0.293 & 0.107 & (0.084 , 0.503) & 0.373 & 0.085 & (0.206 , 0.541) \\
$\beta_{12}$ &  0.139 & 0.105 & (-0.066 , 0.345) & 0.185 & 0.105 & (-0.021 , 0.391) \\
$\beta_{13}$ &  0.363 & 0.070 & (0.227 , 0.500) & 0.364 & 0.075 & (0.218 , 0.510) \\
$\beta_{14}$ &  0.351 & 0.097 & (0.161 , 0.542) & 0.410 & 0.098 & (0.218 , 0.601) \\
$\phi_1$      & -0.087 & 0.085 & & -0.077 & 0.089 \\
$\sigma^2$    &  0.176 & 0.046 & & 0.254 & 0.030 \\
$\nu$         &  5.002 & 3.059 & & - & - \\
\midrule
MSPE & \multicolumn{3}{c}{0.101} &  \multicolumn{3}{c}{0.126} \\
MAPE & \multicolumn{3}{c}{0.240}  & \multicolumn{3}{c}{0.255} \\
\bottomrule
\end{tabular*}
\end{table}

For comparison purposes, we considered the model in Equation \ref{app2} with regression errors following an autoregressive model of order 1 (as defined by Equation \ref{error1}), with innovations $\eta_t$ being independent and identically distributed as $\mathcal{N}(0,\sigma^2)$ and $t(0,\sigma^2,\nu)$ (CAR(1) and CAR$t$(1), respectively). Both models were fitted using the \textsf{R} library \textbf{ARCensReg} through the functions \texttt{ARCensReg} and \texttt{ARtCensReg}, respectively. Parameter estimates and their corresponding standard errors (SEs) are displayed in Table \ref{app2_est}, along with the MSPE and the mean absolute prediction error (MAPE) were computed considering one-step-ahead predictions for the testing dataset. These criteria indicate that the heavy-tailed Student-$t$ ($\widehat{\nu} = 5.002$) model provides better predictions than the normal model for the phosphorus concentration data.

\begin{figure}[t]
\centering
\caption{\textbf{Phosphorus concentration data}. Plots of residuals for the CAR(1) and CAR$t$(1) models.} \label{app2_residual}
\text{CAR$t$(1)}\\
\includegraphics[scale=.68]{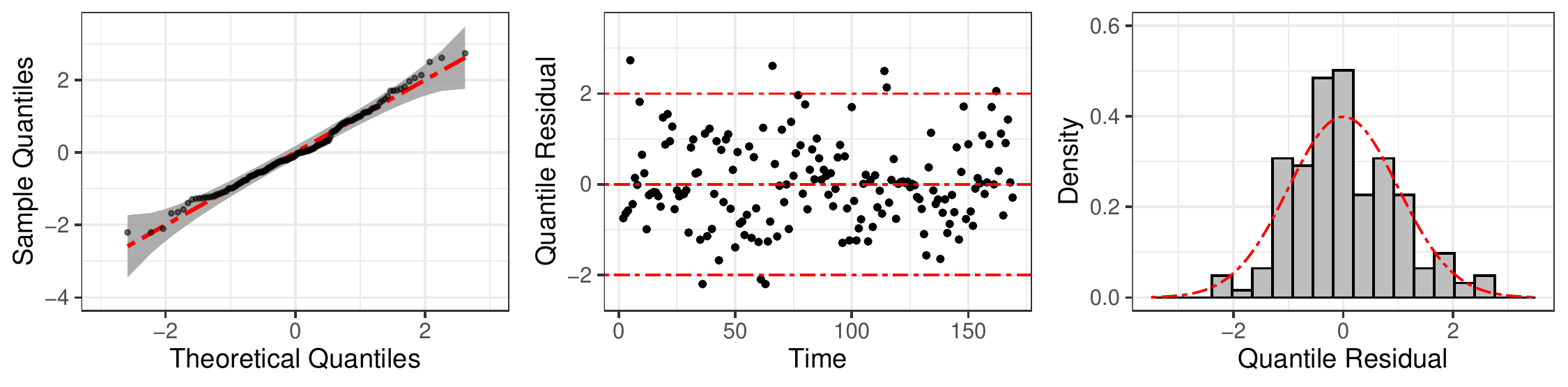}
\text{CAR(1)}\\
\includegraphics[scale=.68]{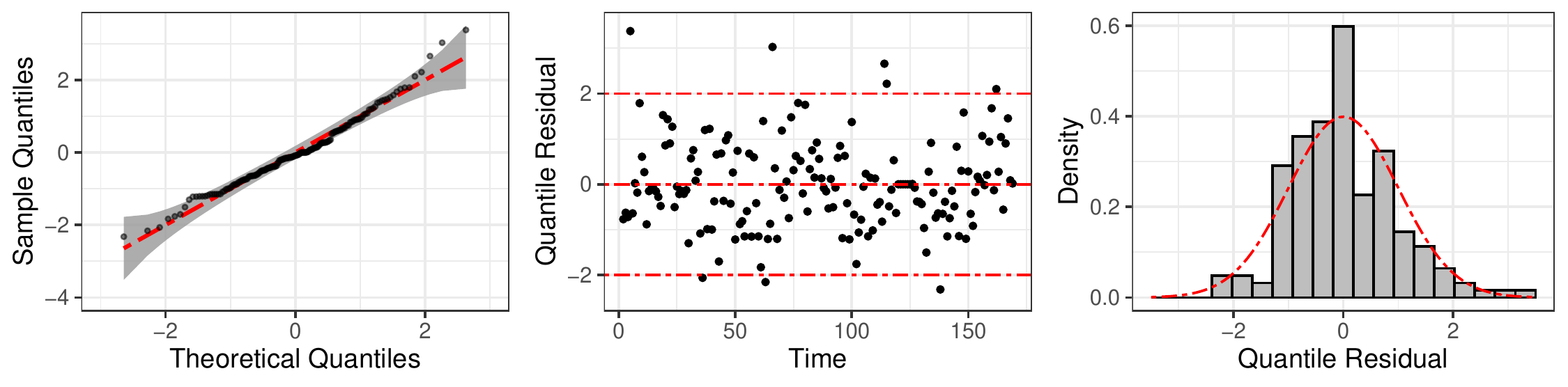}
\end{figure}

One can also note that the estimates for $\beta_{0j}, j=1,2,3,4$ under the CAR$t$(1) model are negative and greater than the estimates obtained from the CAR(1) model. On the other hand, estimates for slopes $\beta_{1j}$ are all positive and, except for the second quarter, they are significantly different from zero, evidencing the correlation between the water discharge and the phosphorus concentration. Regarding the autoregressive coefficient $\phi_1$, both models estimated similar values. 

Figure \ref{app2_residual} presents a Q-Q plot (left), a residual vs. time scatterplot (middle), and a histogram (right) for residual analysis for both models. For the CAR$t$(1) model (top), we see in the Q-Q plot that all points are forming a roughly straight line and lie within the confidence bands. Further, the histogram seems to correspond to a normally distributed variable, and the dispersion plot seems to be related to independent residuals. On the other hand, the Q-Q plot for the CAR(1) model (bottom) presents some points outside of the confidence bands on the upper tail, indicating that the distribution is skewed or heavy-tailed. Additionally, we see from the scatterplot and histogram larger residual values than those from the $t$ model. Therefore, the CAR$t$(1) model seems to provide a better fit to the phosphorus concentration data than the CAR(1) model. Plots with the sample autocorrelation for the quantile residuals are displayed in Appendix \ref{sec:ap:fig}, Figure \ref{acf:app}.

\begin{figure}[t]
\centering
\caption{\textbf{Phosphorus concentration data}. Observed (black solid line) and predicted values considering Student-$t$ (pink line) and normal (light blue line) innovations. Black dashed lines represent the period with missing observations.} \label{app2_pred}
\includegraphics[scale=.7]{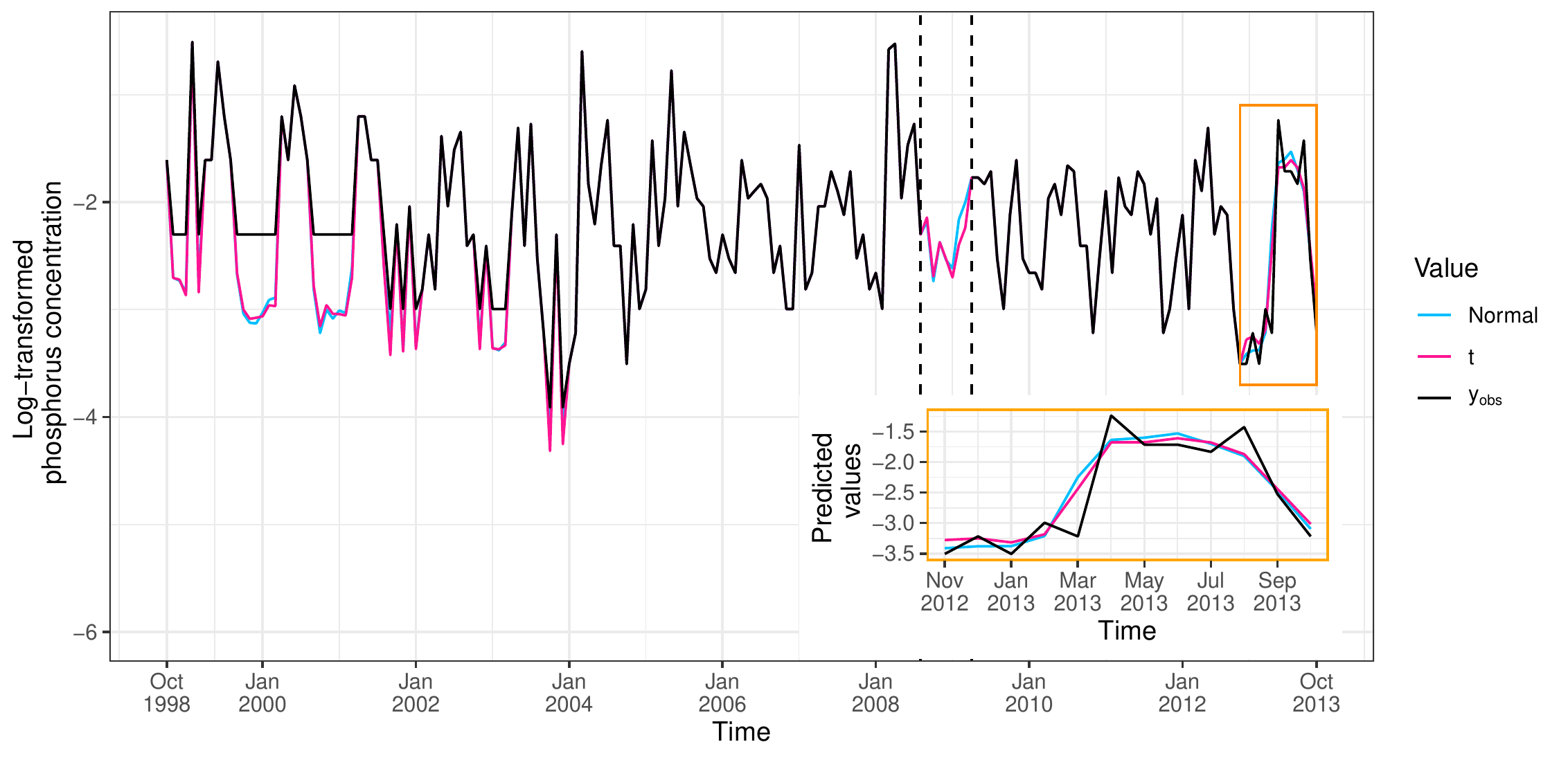}
\end{figure}

Finally, Figure \ref{app2_pred} shows the imputed values for the censored and missing observations from October 1998 to October 2012 (training dataset) and the predicted values from November 2012 to October 2013 (yellow box) considering normal (light blue line) and Student-$t$ innovations (pink line), with observed values represented as a black solid line and vertical black dashed lines indicating the period with missing values. Here, we see slight differences between the predicted values obtained under both fitted models. Besides, the general behavior of the imputed values for the missing period seems to capture well the seasonal behavior of the time series and is also similar for both models. In addition, for assessing the convergence of SAEM parameter estimates, convergence plots for our proposal are displayed in Figure \ref{app2_conver} in Section \ref{sec:ap:fig} of the Appendix.

%%%%%%%%%%%%%%%%%%%%%%%%%%%%%%%%%%%%%%

\section{Conclusions}\label{conclusions}

Extending autoregressive regression methods to include censored response variables is a promising area of research. This paper introduces a novel model that can handle left, right, or interval censoring time series, while simultaneously modeling heavy-tails and missing observations, which can be treated as interval-censored observations.

Our approach extends some previous works, such as \cite{schumacher2017censored} and \cite{liu2019parameter}, and handles missing data while respecting the nature of the data without the need for transformations. 
%Unlike the last work, which can only handle missing data, our proposal can consider responses that take values in specific real subsets. For example, if the response is a proportion or any positive quantity, it is enough to consider the censorship intervals $[0,1]$ and $[0,\infty)$, respectively. In other words, to respect the nature of the missing data without the need for transformations. 
The proposed methods have been coded and implemented in the \textsf{R} package {\bf ARCensReg}, which is available for {the users} on the CRAN repository.

It is important to remark that we assumed the dropout/censoring mechanism to be missing at random (MAR) \cite[see][p 283]{diggle2002analysis}. However, in the case where MAR with ignorability is not realistic, the relationship between the unobserved measurements and the censoring process should be further investigated.
Future directions point to tackling the limitation of assuming that the first $p$ observations are fully observed to fit a CAR$t(p)$ model. Furthermore, a natural and interesting path for future research is to extend this model to a multivariate framework. %An in-depth investigation of such extension is beyond the scope of the present work, but it is an interesting topic for further research.

\section*{Acknowledgments}
This study was financed by the Coordenação de Aperfeiçoamento de Pessoal de Nível Superior - Brasil (CAPES) - Finance Code 001. Larissa A. Matos acknowledges support from FAPESP-Brazil (Grant 2020/16713-0).

%%%%%%%%%%%%%%%%%%%%%%%%%%%%%%%%%%%%%%
%\bibliographystyle{unsrt}  
\bibliographystyle{chicago}  
\bibliography{bibliografia}

%%%%%%%%%%%%%%%%%%%%%%%%%%%%%%%%%%%%%%

\newpage
\appendix

\section{Conditional distributions and derivatives}

\subsection{Full conditional distributions} \label{conditional}
The full conditional distributions are needed to perform the E-Step of the SAEM algorithm; these are $f(\au|\y_o,\y_m,\btheta)$ and $f(\y_m|\au,\y_o,\btheta)$. We have that, the conditional probability density function (pdf) of $\au=(u_1,\ldots,u_n)^\top$ is given by
\begin{eqnarray*}
f(\au|\y_o,\y_m,\btheta) &=& \frac{f(\au,\y_o,\y_m|\btheta)}{f(\y_o,\y_m|\btheta)} \propto f(\au,\y_o,\y_m|\btheta) \\
&\propto&  \prod_{i=p+1}^n u_i^{(\nu-1)/2} \exp\Big\{ -\frac{u_i}{2\sigma^2}\left(y_i-\x_i^\top\bbeta - \y_{(i,p)}^\top\bphi + \bbeta^\top\X_{(i,p)}^\top\bphi \right)^2 - \frac{\nu}{2} u_i \Big\} \\
&=& \prod_{i=p+1}^n u_i^{(\nu-1)/2} \exp\left\lbrace -\frac{u_i}{2}\left(\nu + \frac{(y_i-\x_i^\top\bbeta - \y_{(i,p)}^\top\bphi + \bbeta^\top\X_{(i,p)}^\top\bphi)^2}{\sigma^2} \right)\right\rbrace.
\end{eqnarray*}
From the last expression, we deduce that $U_i$ is independent of $U_j$ for all $i\neq j$, where $U_i|\y_o,\y_m,\btheta \stackrel{ind}{\sim}\G\left(a_i,b_i\right)$, with $a_i=\displaystyle\frac{\nu + 1}{2}$, $b_i=\displaystyle\frac{1}{2}\left(\nu + \frac{\varrho_i^2}{\sigma^2}\right)$, and $\varrho_i=y_i-\x_i^\top\bbeta - \y_{(i,p)}^\top\bphi + \bbeta^\top\X_{(i,p)}^\top\bphi$, for $i= p+1, \ldots, n$.

To compute the condition distribution of $\y_m$, we consider that the first $p$ observations are completely observed. Using the VAR(1) model representation, as \cite{zhou2020student} suggested, we get the expression,
$$\w_t = \bPhi{\w}_{t-1} + \balpha_t,$$ 
where $\balpha_t = (\eta_t, 0, \ldots, 0)^\top$ is a vector of dimension $p$, $\bPhi = [\bphi, \textbf{A}^\top]^\top$ is a $p\times p$ matrix in which $\textbf{A}$ is a $(p-1) \times p$ matrix with the identity matrix in its first $p-1$ columns and $0$s in the last one, $\w_{t} = (\tilde{y}_t, \tilde{y}_{t-1}, \ldots, \tilde{y}_{t-p+1})^\top$, and $\tilde{y}_t = y_t - \x_t^\top\bbeta$. Through a recursive process based on VAR(1) form, we have 
\begin{eqnarray*}
t=p+1 &\Rightarrow & {\w}_{p+1} = \bPhi{\w}_{p} + \balpha_{p+1}. \\
t=p+2 &\Rightarrow & {\w}_{p+2} = \bPhi{\w}_{p+1} + \balpha_{p+2} = \bPhi(\bPhi{\w}_{p} + \balpha_{p+1}) + \balpha_{p+2} = \bPhi^2{\w}_p + \bPhi\balpha_{p+1} + \balpha_{p+2}. \\ 
t=p+3 &\Rightarrow & {\w}_{p+3} = \bPhi{\w}_{p+2} + \balpha_{p+3} = \bPhi(\bPhi^2{\w}_p + \bPhi\balpha_{p+1} + \balpha_{p+2}) + \balpha_{p+3} \\
&& \hspace*{1cm} = \bPhi^3{\w}_p + \bPhi^2 \balpha_{p+1} + \bPhi\balpha_{p+2} + \balpha_{p+3}. \\
&\vdots& \\
t=p+k &\Rightarrow & {\w}_{p+k} = \bPhi{\w}_{p+k-1} + \balpha_{p+k} = \bPhi^k {\w}_{p} + \sum_{j=0}^{k-1} \bPhi^j \balpha_{p+k-j}.
\end{eqnarray*}
Note that the model defined in Equations \ref{modelo1} and \ref{error1} can be recovered through the first element of the preceding vectors, i.e.,
\begin{eqnarray}\label{ar1model}
y_{p+k} = \x_{p+k}^\top\bbeta + \left(\bPhi^k\right)^\top_{1.} ({\y}_{(p+1,p)} - \X_{(p+1,p)}\bbeta) + \sum_{j=0}^{k-1} \left(\bPhi^j\right)_{11}\eta_{p+k-j},
\end{eqnarray}
where $\bPhi^k$ represents the matrix $\bPhi$ multiplied by itself $k$ times, $\big(\bPhi^k\big)_{1.}$ is a $p\times 1$ vector whose elements correspond to the first row of $\bPhi^k$, and $\big(\bPhi^k\big)_{11}$ is the element (1,1) of $\bPhi^k$. From Equation \ref{ar1model}, we deduce that $Y_{p+k}$ given $\y_{(p+1,p)}$, $\btheta$, and $\textbf{U} = \au$ is normally distributed, for all $k=1,\ldots, n-p$. Thus, the conditional expectation and the elements of the variance-covariance matrix that characterizes the normal distribution will be given by
\begin{eqnarray}
\tilde{\mu}_{(k)} &=& \E[Y_{p+k}| \au, \y_{(p+1,p)}, \btheta] \nonumber\\
&=& \E\Bigg[ \x_{p+k}^\top\bbeta + \big(\bPhi^k\big)^\top_{1.} ({\y}_{(p+1,p)} - \X_{(p+1,p)}\bbeta) + \sum_{j=0}^{k-1} \big(\bPhi^j\big)_{11}\eta_{p+k-j} \bigg| \au, \y_{(p+1,p)}, \btheta \Bigg] \nonumber\\
&=& \x_{p+k}^\top\bbeta + \big(\bPhi^k\big)^\top_{1.} ({\y}_{(p+1,p)} - \X_{(p+1,p)}\bbeta). \label{media1} \hspace*{5cm} \\
\tilde{\sigma}_{(kl)} &=& \Cov(Y_{p+k}, Y_{p+l} | \au,\y_{(p+1,p)},\btheta) = \Cov\left(\sum_{j=0}^{k-1} \left(\bPhi^j\right)_{11}\eta_{p+k-j}, \sum_{i=0}^{l-1} \left(\bPhi^i\right)_{11}\eta_{p+l-i} \Big|\au, \y_{(p+1,p)},\btheta\right) \nonumber \\
&=& \sum_{j=0}^{k-1} \sum_{i=0}^{l-1} \left(\bPhi^j\right)_{11} \left(\bPhi^i\right)_{11} \Cov\left( \eta_{p+k-j}, \eta_{p+l-i} \big|\au,\y_{1:p},\btheta \right) = 
\sum_{j=0}^{r-1} 
\frac{\sigma^2}{u_{p+r-j}}
\left(\bPhi^j\right)_{11} \Big(\bPhi^{|k-l|+j}\Big)_{11} \nonumber \\
&=& \sum_{j=1}^{r} 
\frac{\sigma^2}{u_{p+j}}
\big(\bPhi^{k-j}\big)_{11} \big(\bPhi^{l-j}\big)_{11},
\label{var1} \hspace*{2.1cm}
\end{eqnarray}
where $r = \min(k,l)$. Now, suppose that $\y$ is partitioned into two vectors, $\y_{1:p}\in\mathbb{R}^p$ containing the first $p$ observed values and $\y_{-p}\in\mathbb{R}^{n-p}$ containing the remaining observations, i.e., $\y = (\y_{1:p}^\top,\y_{-p}^\top)^\top$. Let $\y_o$ and $\y_m$ be the observed and the censored/missing part of $\y_{-p}$, respectively, then the distribution of $\y_{-p}|\au,\y_{1:p},\btheta \equiv \y_o,\y_m|\au,\y_{1:p},\btheta \sim \N_{n-p}(\tilde{\bmu},\tilde{\bSigma})$, where the $i$th element of $\tilde{\bmu}$ is $\tilde{\mu}_{(i)}$ and the element $(i,j)$ of $\tilde{\bSigma}$ is equal to $\tilde{\sigma}_{(ij)}$, for all $i,j=1,\ldots,n-p$.

To compute the conditional distribution of $\y_m|\au,\y_o,\y_{1:p},\btheta$, we rearrange the elements of $\y_{-p}, \tilde{\bmu}$, and $\tilde{\bSigma}$ as follows
\begin{equation*}
\y_{-p} = \left(\begin{array}{c} \y_o \\ \y_m \end{array}\right), \quad \tilde{\bmu} = \left(\begin{array}{c} \tilde{\bmu}_o \\ \tilde{\bmu}_m \end{array}\right), \quad \mbox{and} \quad \tilde{\bSigma} = \left(\begin{array}{cc} \tilde{\bSigma}_{oo} & \tilde{\bSigma}_{om} \\ \tilde{\bSigma}_{mo} & \tilde{\bSigma}_{mm} \end{array}\right).
\end{equation*}
Using the results for the conditional distribution of a normal distribution, we obtain $\y_m|\au,\y_o,\y_{1:p},\btheta\sim \N(\tilde{\bmu}^*, \tilde{\bSigma}^*)$, with $\tilde{\bmu}^* = \tilde{\bmu}_m + \tilde{\bSigma}_{mo} \tilde{\bSigma}_{oo}^{-1}(\y_o - \tilde{\mu}_o)$ and $\tilde{\bSigma}^* = \tilde{\bSigma}_{mm} - \tilde{\bSigma}_{mo}\tilde{\bSigma}_{oo}^{-1}\tilde{\bSigma}_{om}$.

\subsection{First and second derivatives of the complete-data log-likelihood function}\label{derivadas}

The complete log-likelihood function for the model defined by Equations \ref{modelo1}-\ref{censored} is given by
\begin{eqnarray*}
\ell_c(\btheta; \y_c) = g(\nu|\au) - \frac{n-p}{2}\log\sigma^2 -\frac{1}{2\sigma^2}\sum_{i=p+1}^n u_i\left(\tilde{y}_i - \w_{i}^\top\bphi \right)^2 + cte,
\end{eqnarray*}
where $ g(\nu|\au) = \frac{n - p}{2}\left(\nu\log\left(\frac{\nu}{2}\right) - 2 \log\Gamma\left(\frac{\nu}{2}\right)\right) + \frac{\nu}{2}\left(\sum_{i=p+1}^n \log u_i - \sum_{i=p+1}^n u_i\right)$, $\tilde{y}_i = y_i - \x_i^\top\bbeta$, and $\w_{i} = \y_{(i,p)} - \X_{(i,p)}\bbeta$.\\

\noindent The first derivative of $\ell_c(\btheta; \y_c)$ with respect to each element of $\btheta = (\bbeta^\top, \bphi^\top, \sigma^2, \nu)$ is:
\begin{flalign*}
& \frac{\partial\ell_c(\btheta; \y_c)}{\partial\nu} =  \frac{n-p}{2}\left(\log\left(\frac{\nu}{2}\right) + 1 - \psi\left(\frac{\nu}{2}\right) \right)+ \frac{1}{2}\left(\sum_{i=p+1}^n \log u_i - \sum_{i=p+1}^n u_i\right), & \\
& \frac{\partial \ell_c(\btheta; \y_c)}{\partial\sigma^2} = -\frac{n-p}{2\sigma^2} + \frac{1}{2\sigma^4}\sum_{i=p+1}^n u_i\left(\tilde{y}_i - \w_{i}^\top\bphi \right)^2, & \\
& \frac{\partial\ell_c(\btheta; \y_c)}{\partial\bphi} = \frac{1}{\sigma^2} \sum_{i=p+1}^n \left[  u_i(y_i - \x_i^\top\bbeta)\w_{i} - u_i\w_{i}\w_{i}^\top\bphi \right], & \\
& \frac{\partial\ell_c(\btheta; \y_c)}{\partial\bbeta} = \frac{1}{\sigma^2}\sum_{i=p+1}^n\left[ u_i(y_i-\y_{(i,p)}^\top\bphi)\balpha_i - u_i\balpha_i\balpha_i^\top \bbeta \right], &
\end{flalign*}
with $\balpha_i = \x_i - \X_{(i,p)}^\top \bphi \,$ and $\psi(x) = \frac{\mbox{d} }{\mbox{d} x} \log \Gamma(x) = \frac{\Gamma'(x)}{\Gamma(x)}$. \\

\noindent Besides, elements of the Hessian matrix are given by:
\begin{flalign*}
& \frac{\partial^2\ell_c(\btheta; \y_c)}{\partial\nu^2} = \frac{n-p}{2}\left(\frac{1}{\nu} - \frac{1}{2}\psi_1 \left(\frac{\nu}{2} \right) \right), & \\
& \frac{\partial^2\ell_c(\btheta; \y_c)}{\partial\sigma^2\partial\nu} = 0, \quad
\frac{\partial^2\ell_c(\btheta; \y_c)}{\partial\bphi\partial\nu} = \textbf{0}, \quad
\frac{\partial^2\ell_c(\btheta; \y_c)}{\partial\bbeta\partial\nu} = \textbf{0}, & \\
& \frac{\partial^2\ell_c(\btheta; \y_c)}{\partial(\sigma^2)^2} = \frac{n-p}{2\sigma^4} - \frac{1}{\sigma^6} \sum_{i=p+1}^n u_i\left(\tilde{y}_i - \w_{i}^\top\bphi\right)^2, & \\
& \frac{\partial^2\ell_c(\btheta; \y_c)}{\partial\bphi\partial\sigma^2} = -\frac{1}{\sigma^4}\sum_{i=p+1}^n \left[ u_i\left(y_i - \x_i^\top\bbeta\right)\w_{i} - u_i\w_{i}\w_{i}^\top\bphi \right], & \\
& \frac{\partial^2\ell_c(\btheta; \y_c)}{\partial\bbeta\partial\sigma^2} = -\frac{1}{\sigma^4}\sum_{i=p+1}^n \left[ u_i\left(y_i - \y_{(i,p)}^\top\bphi\right)\balpha_i - u_i\balpha_i\balpha_i^\top \bbeta \right], & \\
& \frac{\partial^2\ell_c(\btheta; \y_c)}{\partial \bphi \partial \bphi^\top} = -\frac{1}{\sigma^2} \sum_{i=p+1}^n u_i\w_{i}\w_{i}^\top, & \\
& \frac{\partial^2\ell_c(\btheta; \y_c)}{\partial\bbeta\partial\bphi^\top} = \frac{1}{\sigma^2}\sum_{i=p+1}^n\left[ u_i\left( \w_{i}^\top\bphi + \x_i^\top\bbeta - y_i \right)\X_{(i,p)}^\top - u_i\balpha_i\w_{i}^\top \right], & \\
& \frac{\partial^2\ell_c(\btheta; \y_c)}{\partial \bbeta \partial \bbeta^\top} = -\frac{1}{\sigma^2} \sum_{i=p+1}^n u_i\balpha_i\balpha_i^\top, &
\end{flalign*}
with $\psi_1(x) = \frac{\mbox{d}^2 }{\mbox{d} x^2} \log \Gamma(x)$.

%%%%%%%%%%%%%%%%%%%%%%%%%%%%%%%%%

\section{Additional results}\label{sec:ap:fig}

This section displays some additional results obtained from the simulation study and the analysis of the phosphorus concentration dataset.

\subsection{Simulation study 1}

\begin{figure}[h]
\centering
\caption{\textbf{Simulation Study 1}. Boxplot of the estimates for $\nu$ in the CAR$t(2)$ model considering different sample sizes and
LOD.} \label{est1:nu}
\includegraphics[scale=.69]{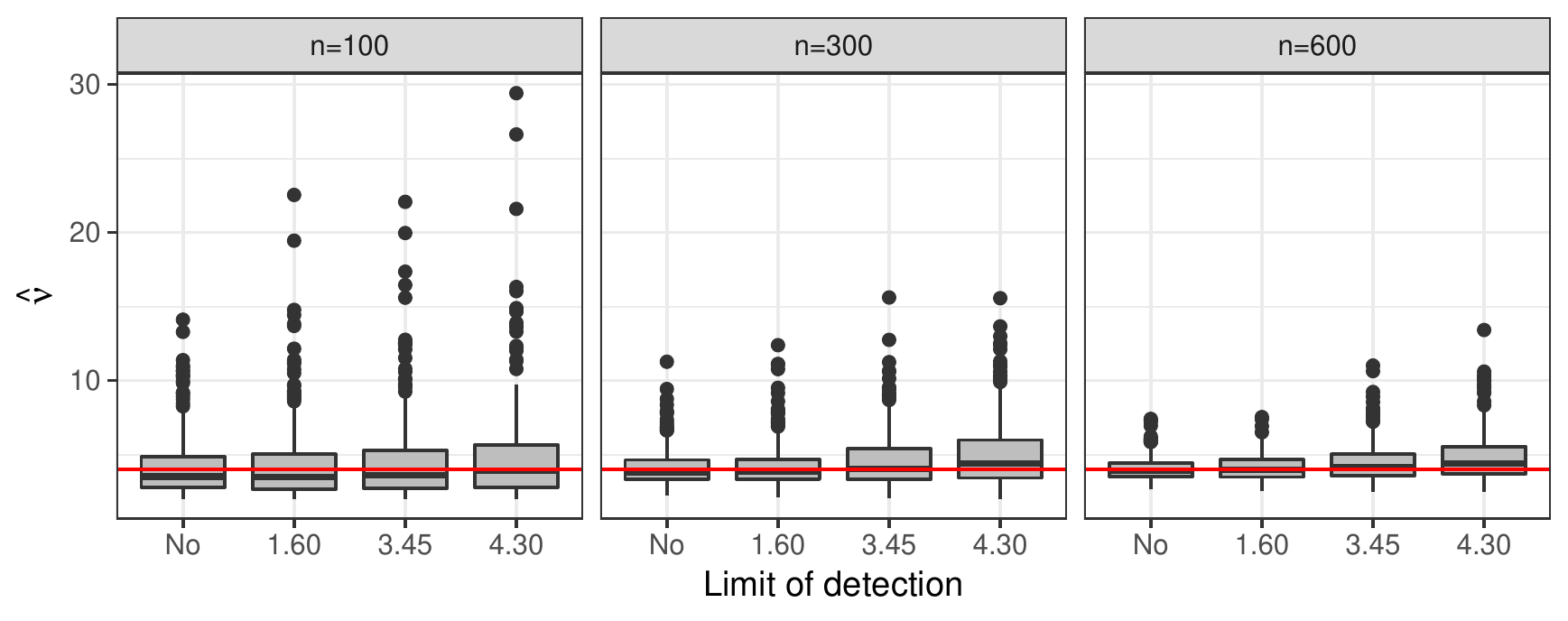}
\end{figure}

Figure \ref{est1:nu} shows boxplots for the estimates of $\nu$ considering different sample sizes and limits of detection (LOD). Here it is possible to observe that the median of the estimates is close to the true value ($\nu = 4$) independent of the sample size and LOD. Furthermore, interquartile ranges decrease as
sample sizes increase, suggesting the consistence of the estimates.

\subsection{Application}

\begin{figure}[!t]
\centering
\caption{\textbf{Phosphorus concentration data}. Sample autocorrelation function for the quantile residuals obtained from the CAR$t$(1) and CAR(1) model.} \label{acf:app}
\includegraphics[scale=.65]{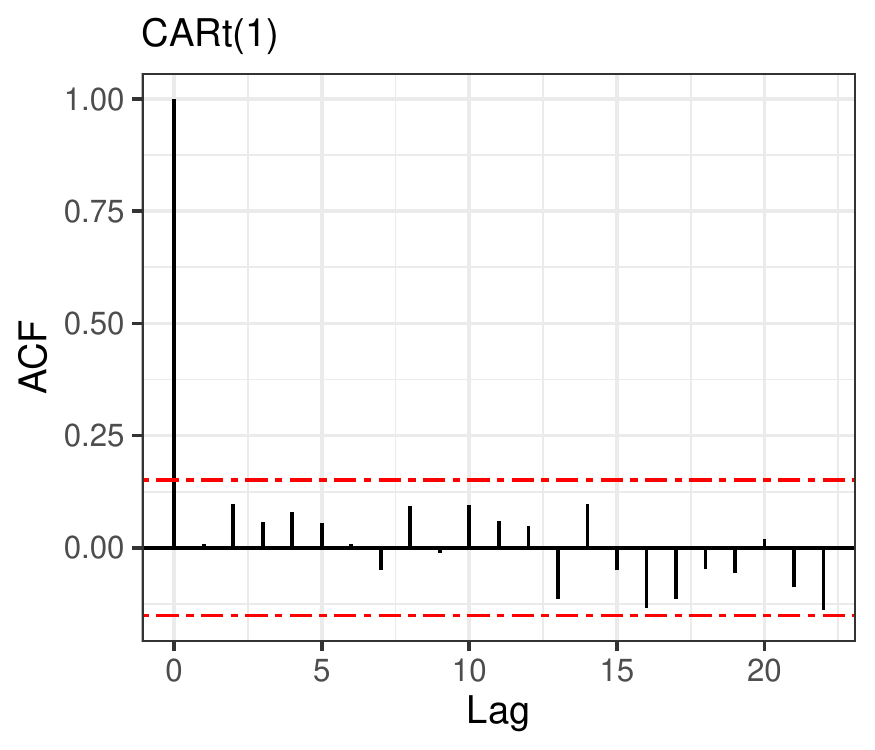}
\includegraphics[scale=.65]{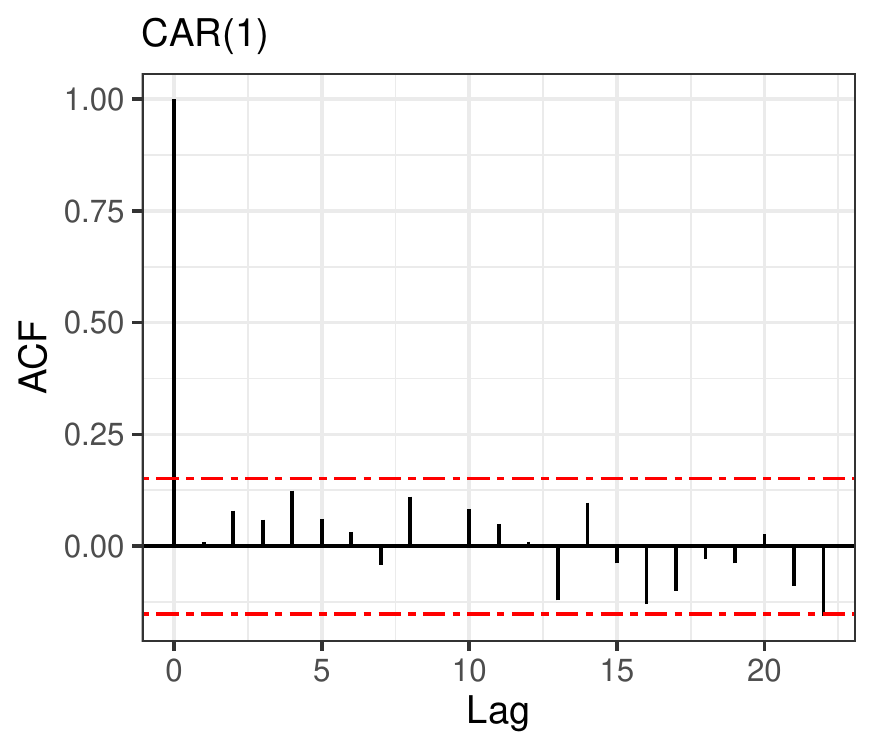}
\end{figure}

For the phosphorus concentration dataset, available in the package \textbf{ARCensReg}, was fitted two censored autoregressive models, one which considered a model with Student-$t$ innovations (CAR$t$(1)) and other with normal innovations (CAR(1)). Then, Figure \ref{acf:app} displays the sample autocorrelation for the quantile residuals computed from each model, where we noted that the obtained residuals present negligible serial autocorrelation.

\begin{figure}[!h]
\centering
\caption{\textbf{Phosphorus concentration data}. Convergence of the SAEM parameter estimates for the CAR$t$(1) model.} \label{app2_conver}
\includegraphics[scale=.7]{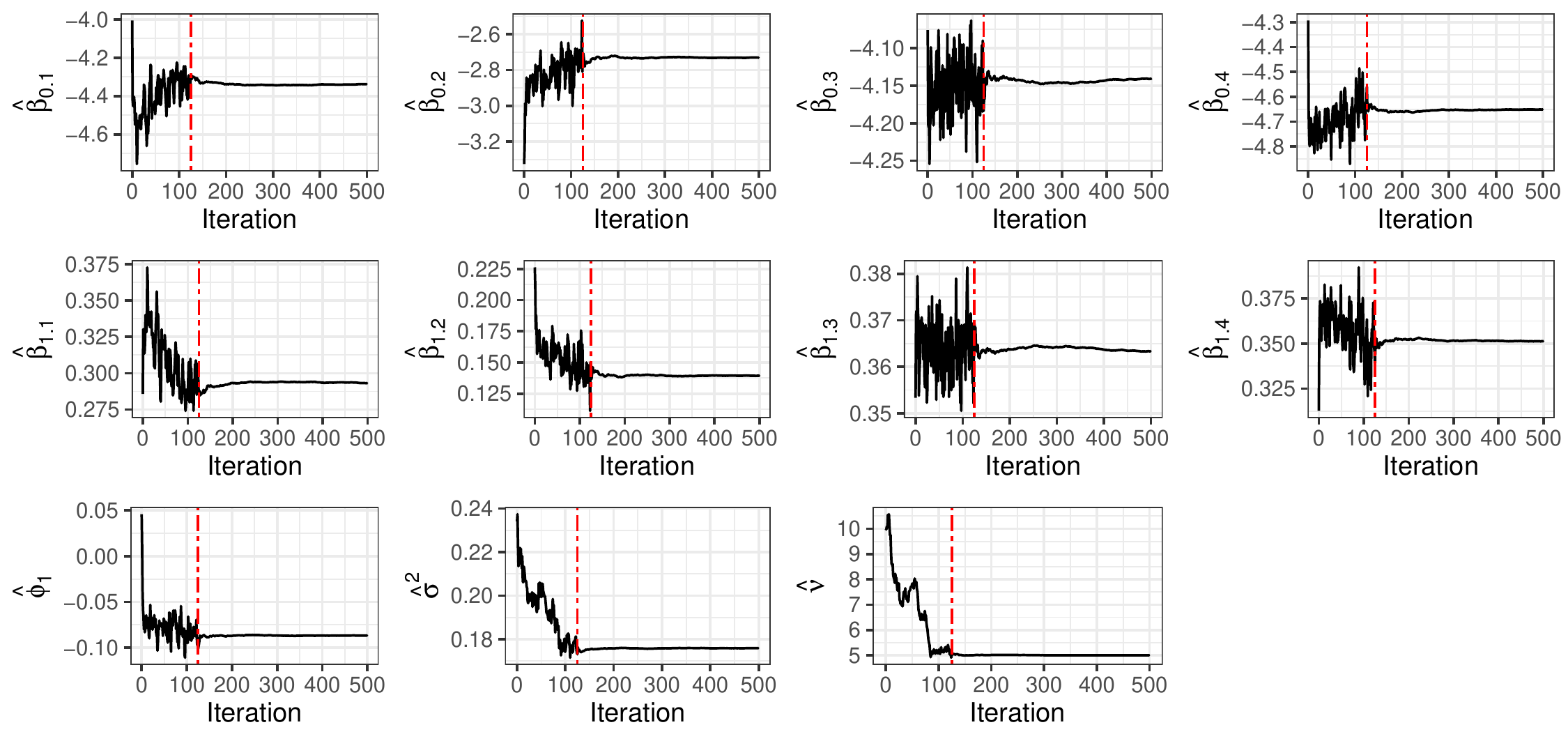}
\end{figure}

Finally, Figure \ref{app2_conver} shows the convergence graphics of the SAEM parameter estimates obtained for the CAR$t$(1) model at each iteration, the model selected as the best for fitting this dataset based on prediction accuracy and residual analysis.

\end{document}